\documentclass[apj]{emulateapj}
\usepackage{lscape}
\def\lsim{\lower.5ex\hbox{$\; \buildrel < \over \sim \;$}}
\def\gsim{\lower.5ex\hbox{$\; \buildrel > \over \sim \;$}}

\def\t{\ifmmode {tau} \else $\tau$ \fi}

\def\ref{\noindent \hangafter=1 \hangindent=0.7 truecm}

\def\cm{\ifmmode {\rm cm}^{-1} \else cm$^{-1}$ \fi}
\def\s{\ifmmode {\rm s}^{-1} \else s$^{-1}$ \fi}
\def\cc{\ifmmode {\rm cm}^{-3} \else cm$^{-3}$ \fi}
\def\cs{\ifmmode {\rm cm}^{-2} \else cm$^{-2}$ \fi}
\def\g{\ifmmode \gamma \else $\gamma$\fi}
\def\G{\ifmmode \Gamma \else $\Gamma$\fi}

\def\kms{\ifmmode {\rm km\ s}^{-1} \else km s$^{-1}$\fi}

\begin{document}

\title{Intrinsic Curvature in the X-ray Spectra of BL Lacertae Objects}

\author{Eric S. Perlman\altaffilmark{1,2}, Greg Madejski\altaffilmark{3,4}, 
Markos Georganopoulos\altaffilmark{1,5}, Karl Andersson\altaffilmark{3,6}, 
Timothy Daugherty$^1$,
Julian H. Krolik$^2$, Travis Rector\altaffilmark{7,8}, 
John T. Stocke\altaffilmark{9}, 
Anuradha Koratkar\altaffilmark{10,11}, Stefan Wagner\altaffilmark{12}, 
Margo Aller\altaffilmark{13}, Hugh Aller$^{13}$, 
Mark G. Allen\altaffilmark{14}}

\altaffiltext{1}{Department of Physics, Joint Center for Astrophysics,
University of Maryland--Baltimore County, 1000 Hilltop Circle, Baltimore, MD
21250, USA.  Email:  perlman@jca.umbc.edu, markos@jca.umbc.edu}

\altaffiltext{2}{Department of Physics and Astronomy, Johns Hopkins University,
3400 North Charles Street, Baltimore, MD 21218, USA}

\altaffiltext{3}{Stanford Linear Accelerator Center, 2575 Sand Hill Road,
Menlo Park, CA 94025, USA}

\altaffiltext{4}{Kavli Institute for Particle Astrophysics and Cosmology, 
Stanford University, Stanford, CA  94305, USA}

\altaffiltext{5}{NASA's Goddard Space Flight Center, Mail Code 660, Greenbelt, 
MD, 20771}

\altaffiltext{6}{Cosmology, Particle Astrophysics \& String Theory (CoPS), 
AlbaNova University Center, Department of Physics, Stockholm University,
Roslagstullsbacken 21,  S-10691, Stockholm, Sweden}

\altaffiltext{7}{National Radio Astronomy Observatory, P. O. Box O,
Socorro, NM 87801, USA}

\altaffiltext{8}{Current Address:  Department of Physics and Astronomy,
University of Alaska -- Anchorage, 3211 Providence Drive, Anchorage, AK 99508,
USA}

\altaffiltext{9}{Center for Astrophysics and Space Astronomy,
University of Colorado, Campus Box 389, Boulder, CO 80309, USA}

\altaffiltext{10}{Space Telescope Science Institute, 3700 San Martin
Drive, Baltimore, MD 21218, USA}

\altaffiltext{11}{Current Address:  Goddard Earth Sciences and Technology
Center, 3.002 South Campus, University of Maryland--Baltimore County, 1000
Hilltop Circle, Baltimore, MD  21250}

\altaffiltext{12}{Landessternwarte Heidelberg, Koenigstuhl, Heidelberg,
69117, Germany}

\altaffiltext{13}{Department of Astronomy, University of Michigan, Ann
Arbor, MI, 48109, USA}

\altaffiltext{14}{Centre de Donnees Astronomique, 11 Rue de
l'Universite, 67000 Strasbourg, France}

\begin{abstract}

We report results from {\it XMM-Newton} observations of thirteen X-ray bright
BL Lacertae objects, selected from the {\it Einstein} Slew Survey sample. The
survey was designed to look for evidence of departures of the  X-ray spectra
from a simple power  law shape (i.e., curvature and/or line features), and to
find objects worthy of deeper  study.  Our data are generally well fit by
power-law models, with three cases having hard ($\Gamma<2; dN/dE  \propto
E^{-\Gamma}$) spectra that indicate synchrotron peaks at $E \gsim 5$ keV. 
Previous data had suggested a presence of absorption features in the X-ray
spectra of some BL Lacs.  In contrast, none of these spectra show convincing
examples of line features, either in absorption or emission, suggesting that
such features are rare amongst BL Lacs, or, more likely, artifacts caused by
instrumental effects.  We find significant evidence for intrinsic curvature
(steepening by $d\Gamma / d({\rm log} E) = 0.4 \pm 0.15$) in fourteen of the
seventeen X-ray spectra.   This cannot be explained satisfactorily via excess
absorption, since the curvature is essentially constant from $0.5-6$ keV, an
observation which is inconsistent with the modest amounts of absorption that
would be required. We use the {\it XMM-Newton} Optical Monitor data with
concurrent radio monitoring to derive broadband spectral energy distributions
and peak frequency estimates.  From these we examine models of synchrotron
emission and model the spectral curvature we see as the result of episodic
particle acceleration.

\end{abstract}


\section{Introduction}

Among all active galactic nuclei, BL Lacertae (BL Lac) objects are the most
dominated by variable, non-thermal emissions.  BL Lacs often vary on time
scales of days to hours at frequencies from optical through the $\gamma$-rays
(see Ulrich, Maraschi \& Urry 1997 for a review). Their emissions are dominated
by a broad, featureless continuum, believed to originate in a relativistic jet
oriented very close to our line of sight (see Urry \& Padovani 1995 for a
review). This component is thought to be responsible for their variability as
well as their high optical polarization (Jannuzi, Smith \& Elston 1994) and
radio core-dominance (Rector et al. 2000, Rector \& Stocke 2001 and references
therein). The spectral energy distributions (SEDs) of BL Lacs appear to be
dominated by synchrotron emission at radio to ultraviolet energies (up to X-ray
energies for X-ray selected objects) and inverse-Compton emission at higher
energies.

Previous observations of the X-ray spectra of BL Lacs have yielded somewhat
confusing results.  
Worrall \& Wilkes (1990) compiled the first examination of
a relatively large sample of BL Lacs with {\it Einstein}.  They found power-law
shapes, with a wide range of spectral indices.  This was confirmed by 
observations with HEAO-1 (Sambruna et al. 1994), ROSAT (Perlman et al. 1996a,
Urry et al. 1996, Sambruna et al. 1996), {\it ASCA} (Kubo et al. 1998) and {\it
BeppoSAX} (Wolter et al. 1998; Beckmann et al. 2002; Padovani et al. 2001,
2004).  Those observations also found a pattern of harder X-ray  spectra in
objects with spectral peaks in the infrared to optical (often  called
low-energy peaked BL Lacs, or LBLs) than in those with spectral peaks in the UV
to X-rays (often called high-energy peaked BL Lacs, or HBLs).  The correlation
between peak frequency and spectral index is strong, but not monotonic
(Padovani, Giommi \& Fiore 1997, Lamer et al. 1996, Padovani et al. 2001).  The 
explanation put forward most often for the complex nature of
this relationship was that it was due to observing the intrinsically curved
synchrotron and inverse-Compton spectral components, peaked at a variety of
frequencies, over only a restricted energy range.

If indeed there is curvature in the X-ray spectra of BL Lacs, one might
expect to see evidence of this when larger bandpasses are examined,
particularly in bright objects where high signal-to-noise can be attained.  
Several authors have found evidence of such curvature (Inoue \& 
Takahara 1996; Takahashi et al. 1996; 
Tavecchio et al. 1998; Massaro et al. 2004a, b; Giommi et al. 2002)
among a significant fraction (up to 50\%) of HBLs. However,
those authors did not analyze in detail the alternative possibility of an
additional absorbing column, as had been assumed by other workers (e.g.,
Perlman et al. 1996a, Urry et al. 1996, Kubo et al. 1998, Wolter et al. 1998,
Beckmann et al. 2002, Padovani et al. 2001).  

Another issue is whether line features are present in the X-ray spectra.  Early
{\it Einstein} grating data for PKS 2155$-$304 suggested a soft X-ray deficit
(Canizares \& Kruper 1984), as did later  {\it Einstein} Solid State
Spectrometer data for a few of the brightest objects   (Urry, Mushotzky \& Holt
1986;   Madejski et al. 1991).  Examination of {\it ASCA} and {\it BBXRT} 
spectra of some of the brightest BL Lacs indicated a similar soft X-ray
deficit,  requiring a recovery of the spectrum towards even lower 
energies,  as
inferred from ROSAT PSPC data  (PKS 2155$-$304, Madejski et al. 1992;
H1426+428, Sambruna et al. 1997; PKS 0548$-$322, Sambruna \& Mushotzky 1998). 
These features were all explained by invoking X-ray absorption features at
0.5--0.8 keV.  However, not all bright  BL Lacs were found to require such
features (e.g., Mrk 421, Guainazzi  et al. 1999); nor were such features found
in stacked {\it ROSAT}  spectra of the fainter EMSS BL Lacs (Perlman  et al.
1996a). This lack of consensus regarding the presence or  lack of X-ray
spectral curvature and line features  motivated us to use {\it XMM-Newton} to
obtain high signal  to noise (S/N) spectra of a significant sample of X-ray
bright BL Lac objects with a single instrument.  

The paper is organized as follows.  In \S 2, we describe the sample,
observations and data reduction procedures.  In \S 3 we give the results of our
X-ray spectral fits, specifically concentrating on  issues regarding spectral
curvature and the presence or lack of spectral lines. In \S 4, we focus on
models for X-ray spectral curvature and particle acceleration. Finally, in \S
5, we conclude with a summary and a discussion of the implications of our
results.

\section {Sample, Data and Data Reduction}

\subsection {Sample Design}

Our targets were selected from the {\it Einstein} Slew Survey sample of BL Lacs
(Perlman et al. 1996b).  The Slew Survey is ideal for this purpose because it
is a nearly all-sky survey, with mean limiting flux
F (0.3--3.5 keV) $\sim 1.4 \times 10^{-11} {\rm ~erg ~cm^{-2} ~s^{-1}}$ (Elvis
et al. 1992).   Besides being the largest collection of BL Lacs at such
high X-ray fluxes, the Slew Survey was the first sample to contain
statistically significant numbers of both HBLs and LBLs.  We used this property
to design a sample of 36 objects, which are the X-ray brightest in the LBL and
HBL  sub-classes. Of these, 11 (nearly all HBLs) were on the {\it XMM-Newton}
or {\it Chandra} observing lists of other projects and so we did not choose to
observe them again.  We were awarded  observing time for the 12 X-ray brightest
of the remaining 24 objects, which unfortunately include only the HBL
subclass:  LBLs were approved only in priority C and none were observed.  
under a separate proposal {by some members of our team.}  Table 1 lists
the objects we observed.  We do not discuss here the objects observed by
other workers, which have already appeared in the literature (e.g.,
Boller et al. 2001; Watson et al. 2004; Blustin, Page \& Branduardi 2004;
Cagnoni et al. 2004); however, where appropriate we make use of 
their results.

\subsection{Instruments and Observations}

We observed each object with all instruments aboard {\it XMM-Newton.}
The {\it XMM-Newton} Observatory (Jansen et al. 2001) consists of three
coaligned 7.5 m focal length X-ray telescopes, focussing X-rays onto the
European Photon Imaging Camera (EPIC) and two Reflection Grating Spectrometers
(RGS1 and RGS2 respectively).  The EPIC has three detectors, sensitive in the
0.3-10 keV band: two metal-oxide-semiconductor (MOS) and one p-n junction (PN)
CCD arrays.  These instruments provide moderate-resolution X-ray imaging (PSF
FWHM $6''$) and spectroscopy  ($E/\delta E \approx 20-50$, depending on
energy). The two RGS  instruments provide spectroscopy at higher resolution, 
with $E/\delta E \approx 300$.  

Because the main goal was to address the sample properties of the BL Lac class,
rather than to obtain very deep observations of individual objects, our
integration times were short, $\sim 5$ ks on each instrument. Table 1 lists the
EPIC PN and MOS good on-source  times and count rates. Figure 1 shows two
examples of our EPIC observations, fit with a single-power-law model (\S 3). As
can be seen,  the S/N of the EPIC data were quite high (20-50 below 1 keV;
lower at high energies).  The RGS spectra (\S 3.3) had much more modest S/N,
due to the lower effective area of the gratings.   In all but one observation,
the EPIC was used in small-window mode in order to minimize the effects of
pileup.  On most objects, the THIN filter was used in both the PN and MOS;
however, for the brighter objects we used the MEDIUM filter to avoid
contamination from optical light due to the object (none of the targets has
other X-ray bright objects within the field).  

Eight {\it XMM-Newton} observations were significantly affected by
proton flares.  Of these, six were impacted badly enough that one or more of
the X-ray instruments had to be shut off because of the high background, and in
two, all instruments were shut off during the observation. For the six
observations where X-ray instruments were shut off completely due to  high
background, the observation was automatically attempted again (see Table 1 for
details).  As a result, four objects were observed twice and one object was
observed three times.   We inspected all data from {\it XMM-Newton}, regardless
of whether the observation was 
affected by flares.  We generally found that even for the
observations impacted by proton flaring, some of the X-ray data were usable as
long as the instruments were not shut off.

{\it XMM-Newton} also has a coaligned 30 cm optical/UV telescope, the Optical
Monitor (OM).  The OM is sensitive between $\lambda=1000-6000$~\AA\ and has a
typical angular resolution of $\sim2''$.  OM data were gathered for all
objects, but in only one band due to observing restrictions in AO1 that did not
allow multi-band observations within $\la$ 2 hours.  All but one OM dataset was
gathered using the UVW2 filter, which is sensitive between $2000-3000$~\AA. 
Data were taken in exposures of 800 s in order to allow for the possibility of
finding short-timescale optical variability, but no such  variations were
found.  We include in Table 2 the near-UV magnitude of each source, as derived
from the OM data.

We also monitored the radio fluxes of all but one of these objects using the
University of Michigan Radio Observatory (UMRAO) 26-m telescope
(Aller et al. 1985).  We
observed each object at one or more of three frequencies:  4.5, 
8.0 and 14.5 GHz, but some were below its sensitivity limitations (typically
50 mJy at 8.0 GHz). These data are also given in Table 2; two example datasets 
are shown in Figure 2.

\subsection{Data Reduction Procedures}

All source and background extraction, as well as examination of the
lightcurves, was done in the {\it XMM-Newton} Science Analysis System (SAS)
v5.4.1.  We performed a standard reduction of the events list for the PN, MOS
and RGS data, which involves the subtraction of hot and dead pixels, removal of
events due to electronic noise, and the correction of event energies for charge
transfer losses.  The PN files were also filtered to include only single,
double, and single+double events (PATTERN $\leq$ 4) while the MOS included all
single to quadruple events (PATTERN $\leq $ 12). Photons were extracted
in all instruments using a $45''$ aperture (which, for the PN and MOS, was
circular in shape).  In  cases where pile-up was significant, an annular
extraction region was used to filter out the  pixels affected most severely.

We extracted lightcurves for every dataset, and filtered out all
high-background time intervals.  To do so, we determined the  count rate in the
10 -- 15 keV band for the entire PN and MOS detectors.  Since the source count
rate in this  band is quite small, intervals with more than 0.3 ct s$^{-1}$ 
for the PN and 1.5 ct s$^{-1}$ for MOS were rejected.   Once this was done, 
the datasets were searched for evidence of source  variability on $<5$ ks
timescales (using the nominal 0.3 -- 10 keV band);  no such variability was
found during any of these observations.  Spectra were extracted, as discussed
in the next section. These spectra were limited to the energy range 0.2--10 keV
for the MOS and 0.15--15 keV for the PN;  however, for the  detailed spectral
fits, we used a more restricted energy range, as given  below, to account for
possible residual calibration uncertainties (see  Andersson \& Madejski 2004
for detailed discsusion).  

For the MOS data, the background region used was the same size as the source
region in a place where there are no visible sources (typically off axis).  For
the PN data, the background region was once again the same size as the source
region at a point where the RAWY values were the same as the source.  Because
of the high fluxes of these sources and the use of the small-window mode  (as
well as the relatively stringent criteria to reject data  segments affected by
flares), we did not extract background maps (see e.g., Read \& Ponman 2003;
note that the background maps vary little within the portion of the detectors
used for small-window mode).  It should, however, be noted that because of
the  use of the small-window mode the background regions are of necessity close
to the source itself, and thus it is possible that some small amount of source 
flux could be subtracted.  This effect causes mainly a small error in the
absolute flux of the sources, and thus does not affect our analysis of possible
curvature or line features.  We did, however, attempt to minimize this effect
by placing the background region as far away from the source as possible within
the small window.  

For the OM data, we performed simple aperture photometry, with a $6''$
circular aperture, using an annulus for background subtraction.  The
size of the annulus was typically $20''$; however, in the case of a
few objects, a nearby companion necessitated some adjustment.  We then
used the zero-points published in the {\it XMM-Newton Observer's
Handbook} (Issue 2.1) to calculate absolute optical fluxes.  
The UMRAO data were reduced  according to the procedure outlined in Aller et
al. (1985).  The fluxes and magnitudes from UMRAO and the OM at the time of
observation are summarized in Table 2.

\section{X-ray Spectral Fits}

As the sources were well centered in the XMM observing region, we used standard
response (RMF) matrices{\footnote{These files are available from
ftp://xmm.vilspa.esa.es/pub/ccf/constituents/extras/responses/}}, picking
the one closest in time to each observation. The Ancillary Response Files
(ARFs) were generated using the arfgen function of SAS 5.4.1.  To assure the
validity of Gaussian statistics,  we grouped the data, combining the
instrumental channels such that each new bin would have at least 40 counts. 
 The exceptions are 1ES 0145+138 (OBSID 0094383401) and 1ES 1255+244 (OBSID
0094383201), where because of the small number of counts we rebinned to 10
counts/bin for the MOS data and 15 counts/bin for the PN data.  All the PN data
were fit in the 1.1--10.0 keV range, while all the MOS data were fit in the
0.3--10.0 keV range, except for  those of the faintest objects which we capped
at 7.0 keV due to low count rates at the highest energies.

All X-ray spectral modeling was done in XSPEC v11.0.  Continuum models were fit
only to the PN and MOS data and then applied to the RGS data.  We used both the
EPIC and RGS data to search for absorption and/or emission lines (\S 3.3).  The
PN and MOS data were fit simultaneously as well as individually  in order to
spot any calibration problems and/or instrument specific problems that might
have occurred during the observation.  Where multiple observations of an object
were obtained (Table 1, \S 2.2), each observation was reduced and analyzed
separately, as we expected that the flux and X-ray spectrum would change
between those observations, as was indeed the case.

\subsection{Spectral Modeling in the broad X-ray band} 

Initially, two models were fit to each EPIC spectrum:  

1. A single power law, 

$dN/dE = k e^{-\sigma(E)N_{H, Gal}}
e^{-\sigma(E)N_{H, int}(1+z)} ~E^{-\Gamma}$.

2. A logarithmic parabola, 

$dN/dE = ke^{-\sigma(E)N_{H, Gal}}
e^{-\sigma(E)N_{H, int}(1+z)} ~E^{(-\Gamma+\beta Log(E))}$.

{\noindent {In each case, we attempted fits with two different treatments of
the absorbing column:  with $N_{H,int}=0$  and with  $N_{H,int}$ (assumed to be
at the redshift of the source) allowed to  vary freely.  We took values of 
$N_{H, Gal}$ from Stark et al. (1992), although it is probable that the actual
Galactic absorbing column for each object might be slightly different due to
the low resolution of the Stark et al. survey.   In all cases, the opacity
associated with the intervening column density was modeled by a standard
Morrison \& McCammonn (1983) absorption model.    The motivation to try both
the power law model, as well as the less-used logarithmic parabola model
(described in Giommi et al. 2002 and references therein), was that both simple
power-law and continuously curving spectral shapes can reasonably be expected
depending on the model adopted for synchrotron aging and acceleration (Leahy
1991, Massaro 2002).    Our spectral fits for the single power-law model are
summarized in Table 3.  In Figure 1, we show as examples spectra of two objects
fit with  the single power-law model.}}

As can be seen, the spectral indices we found range from $\Gamma=1.7-2.9$
(corresponding to  energy spectral indices $\alpha=0.7-1.9$), with twelve of
seventeen  being in the range $\Gamma=2.1-2.7$.  These spectral indices are
similar to the findings of past observations of HBLs by ROSAT (Perlman al.
1996a, Padovani et al. 1997), {\it ASCA} (Kubo et al. 1998) and {\it
BeppoSAX} (Wolter et al. 1998, Beckmann et al. 2001). Three spectra were found
to be flat ($\Gamma<2$).  These objects are likely to have synchrotron peaks at
energies higher than $\sim 5$ keV (see also \S 4).

In the three objects where repeated observations were made, we see significant
variability both in flux and spectral shape (Table 3).  In all of those
objects, the highest flux state observed also has the hardest spectrum, in
agreement with the known spectral variability properties of BL Lac objects
(e.g., Ulrich et al. 1997 and references therein). However,
the correlation between a harder  spectrum and higher X-ray flux is not
one-to-one, as indicated by the three observations of 1ES1959+650. 

\subsection{X-ray Spectral Curvature}

In fourteen of seventeen observations, a better fit (significant at the $>99$\%
level according to $F$-test results; see Table 3) was obtained by allowing for
the possibility of spectral curvature (over and above the default,
power-law plus Galactic $N_H$ model).  This curvature can either be intrinsic,
or the result of additional absorption due to material either within the BL
Lac's host galaxy or our own.  As indicated above, we first investigated the
possibility that additional absorption was present; the result of these
procedures is given in Table 3.  As can be seen, in 13 of 14 observations the
column required is $\gsim 20$\% of the Galactic figure.  We also investigated
the alternate possibility of  intrinsic curvature, using the logarithmic
parabola model as well as an alternate model, detailed later,  where we
simply fitted power-laws independently in four sub-bands,  and obtained 
results of similar significance.   The results of this procedure are given  in
Table 4 and discussed below. Importantly, it is impossible to discriminate {\it
a priori}  between the two models due to their similar mathematical forms (see
above).   It is, however, possible to use other information  and perform other
tests.

We first inspected the implied columns to check the reasonability of the idea
that the spectral curvature was the result of additional absorbing material. We
believe that additional absorbing material is less likely for three reasons. 
First, the typical BL Lac's host galaxy is a bright elliptical, where columns
this high would not normally be expected except in extreme cases (Goudfrooij et
al. 1994). Also, in 1ES1959+650, one of the three objects where repeated
observations were done, the implied $N_H$ appears to vary between epochs.  This
requires  the absorbing material to have been in a region smaller than  $\sim
1$  light-month in size.  This would imply densities $n \gsim 10^3-10^4 {\rm
~cm^{-3}}$, which is outwardly not unreasonable for the inner regions of the
AGN.  But the implied density would have to climb with decreasing timescale,
as  $\tau^{-3/2}$, meaning that unreasonable column densities  would be
reached  if $N_H$ appeared to vary on timescales only 1-2 orders of magnitude
smaller, which are not uncommon for flux variability among BL Lacs (Ulrich et
al. 1997). Finally, there is no indication from optical or UV spectroscopy that
BL Lac objects would have any significant neutral material in the line of sight
to the nucleus (e.g., Perlman et al. 1996; Rector et al. 2000; Rector \& Stocke
2001; Kinney et al. 1991; Lanzetta, Turnshek \& Sandoval 1993; Penton \& Shull
1996).  This material, in principle, can be  partially ionized, providing no
opacity in the optical and UV, but  absorbing via individual edges in the soft
X-ray band.  This can be ruled  out as a general property  on the basis of an
absence of any individual spectral features in the  RGS data (\S 3.3), although
we cannot exclude it in the spectra of fainter objects.  

This is persuasive but not conclusive evidence that these spectra have 
intrinsic X-ray spectral curvature.  To test this idea further, we  split the
{\it XMM-Newton} data into four sub-bands:  0.5--1 keV, 1--2 keV, 2--4 keV and
4--10 keV, and fit each sub-band individually with a power-law model.  For this
exercise we assumed an $N_H$ value fixed at Galactic, and used the MOS
data  only below 2 keV and both the PN and MOS data at higher energies.  This
was  done because of the known issues in cross-calibrating PN data with those
from  the MOS at lower energies, especially near 1 keV (see also the first
paragraph of \S 3, above).  The results of this procedure are given in Table
4, and four examples of these fits are shown in Figure 3.   We note that
the overall $\chi^2$ values for these fits are comparable to those achieved by
varying $N_H$; however it is impossible to  use an $F$-test to compare them
because of the differing ways in which data  was selected for this procedure.
As can be seen, the typical X-ray  curvature seen in our sample is nearly
constant, even up to the highest energy band.  We have characterized this
curvature by looking at the $d\Gamma/d(\log E)$ for each object.  This
information is included in Table 4.  As can be seen, our data are consistent
with a $\langle d\Gamma/d(\log E)\rangle \approx 0.4 \pm 0.15$ in fourteen  of
seventeen  observations (the same ones for which  spectral curvature was
indicated as mentioned above).   This is not consistent with the
observed columns in these objects, which range from  $3-20 \times 10^{20} {\rm
~cm^{-2}}$.    At such moderate columns, it is typical for the spectrum
above about 1 keV to be relatively insensitive to the absorbing column.  This
is yet another argument (albeit indirect) against the presence of additional 
absorbing material explaining the observed spectral curvature in these objects.

As already noted, we also attempted to model intrinsic curvature by
fitting a logarithmic parabola form.  The information for these fits is also
given in Table 4. The logarithmic parabola is characterized by both a spectral
index and a curvature parameter, $\beta$.  The spectral form is somewhat
different mathematically from either of the other forms; however, it is
essentially similar to the application of a constant $d\Gamma/d(\log E)$, with 
the added advantage of allowing a consistent fit in the entire band. In theory,
the value of the curvature parameter $\beta$ should be equivalent to the value
of  $\langle d\Gamma/d(\log E) \rangle$  found using several  sub-bands. 
Inspection of Table 4 shows that these are indeed similar, albeit with small
variations in each case.  The mean curvature required for the objects where
this is indicated by the $F$-tests is essentially identical, in fact:  
$\langle \beta \rangle = 0.37$ versus $\langle d/\Gamma/d(\log E) \rangle =
0.38$.

Giommi et al. (2002) also found good evidence for intrinsic X-ray
spectral curvature (which they modeled by the same logarithmic parabola form)
in about half of the HBLs observed by {\it BeppoSAX}.  The value of $\langle
\beta \rangle$ they found is consistent with our  findings.  However, the
proportion of  curved spectra in Giommi et al. (2002) is considerably lower
than we see.  The difference cannot be explained by our small sample, as twelve
of thirteen objects observed by BeppoSAX were also observed by us with {\it
XMM-Newton}, and of those, for six the best representation of their {\it
BeppoSAX} spectrum was a logarithmic parabola plus fixed absorption (although
n.b., Giommi et al. 2002 did not fit a variable absorption model for any
object).  We believe it is much more likely that the reason for the larger
fraction found here lies in the higher signal to noise and resolution of these
spectra compared to those from {\it BeppoSAX}.  This is supported by the fact
that if one examines the variable $N_H$ fits, the 6 objects where Giommi et al.
(2002) required a logarithmic parabola are best fitted by significantly higher
absorbing columns than those which Giommi et al. (2002) was able to fit
adequately with a single power law.

Our conclusion from the above is that we believe the most appropriate 
model for the spectral curvature we see is intrinsic, rather than due to
additional absorption.  Given the similar values of $\chi^2$, our primary
motivation for concluding this is the indirect lines of evidence presented
above regarding the implausibility of  additional absorbing columns under these
circumstances.  We will proceed on this basis in the upcoming discussion. 

To close this discussion of intrinsically curved spectra, it is useful to
compare the fits and shape of the spectrum achieved for a  given object under a
simple power law model (with Galactic $N_H$) with those implied by the 
logarithmic power law form.  We show this comparison in Figure 4, for one of
the spectra of 1ES1959+650, specifically OBSID 0094383301. There comparison is
shown in two forms: at top, we show the comparison in terms of the detected
spectrum (folded into the instrumental response), while at bottom it is shown
in terms of emitted energy.  The plots at top enable us to see the real
differences in the quality of the fit:  the simple power law model (Figure 4,
{\it top left}) clearly overpredicts the observed spectrum at $E>1$ keV, and
slightly underpredicts the lower-noise MOS data at 1-1.5 keV.  But at higher
energies it is the PN data which have the greater ability to distinguish
between the two models, and the simple power law model overpredicts the data at
energies higher than about 5 keV.  As can be seen, the logarithmic power law
model (Figure 4, {\it top right}) fits the data much better.  The plots at
bottom enable us to see the form of the curvature:  note that above $\sim 1$
keV essentially no curvature would be seen, if the spectrum followed a simple
power law model (Figure 4, {\it bottom left}).  However, under  the logarithmic
parabola model (Figure 4, {\it bottom right}), we see that there is in fact
significant  curvature seen at higher energies, albeit of a mild, gradual form.
This is presumably why this curvature was not seen by earlier satellites, which
suffered from either a much smaller spectral range (for all satellites except
{\it BeppoSAX}) and/or much lower signal to noise (all other satellites).

\subsection{Spectral Line Features in BL Lacs?}

We inspected each spectrum thoroughly for spectral features.  We did not find
convincing evidence for line features in any of these spectra. To quantify
this, both the EPIC and RGS data for the 12 highest signal-to-noise spectra
were searched for absorption features in XSPEC (we did not search datasets
where either instruments were turned off and/or the statistics were
inadequate). The best-fit broadband model (Table 4) was used as the seed model
in each case, and the RGS data were grouped in bins with a minimum of 20
photons apiece (following our practice for the EPIC data; \S 3.1).  We then
artificially set a line feature of 5 eV physical width, and covering fraction
of 0.5, and then stepped it through the range 0.5 to 4.5 keV in increments of
0.01 keV using STEPPAR, and assessed the significance of features using the
$\Delta \chi^2$ (A $3\sigma$ feature has $\Delta \chi^2 = -9.21$, while a $4
\sigma$ feature has $\Delta \chi^2 = -18.42$).   Higher energies were not
searched due to the small effective area and poor statistics.   Once features
were found, we then fixed the energy and allowed XSPEC to converge on a
best-fit width.  This procedure found six narrow features, as shown in Table 5,
in four of the datasets, but in most (1ES 0120+340 OBSID 0094382101, 1ES0323
OBSID 0094382501, 1ES 0414+009 OBSID 0094383101, 1ES0647+250 OBSID 0094380901,
1ES1133+704 OBSID0094170101, 1ES1255+244 OBSID 0094383001, 1ES 1553+113 OBSID
0094380801, 1ES1959+650 OBSID 0094383501), no features were found. 

Are these features real?  As can be seen, all but one are between
3-4 $\sigma$, with one barely over 4$\sigma$.   For Gaussian statistics, 
$P(3\sigma) = 0.0027$, while $P(4\sigma)=0.000318$.    As can be seen
from Table 3, the typical EPIC dataset in our sample is $\sim 300$ channels, as
is the typical RGS dataset (n.b., the number of channels is not significantly
greater due to the relatively poor statistics resulting from the short exposure
times).  Thus the result of  5 features at $\gsim 3 \sigma$  and 1 feature at
$>4 \sigma$ in $\sim 3600$ spectral
channels is consistent with what one would expect from statistical noise 
(respectively $10 \pm 3$ features at $>3 \sigma$ and $1 \pm 1$ feature at $>4
\sigma$).
Moreover, as noted in Table 5, none of the features found in any of the spectra
are present in the spectra from all the EPIC and RGS instruments.  Thus, we
believe all are the result of random noise and therefore not real spectral
lines.  The limits shown in Table 5 can be taken as representative of the
detection limits of our data, which range from $\sim 5$ up to $\sim 50$ eV
depending on count rate.

This conclusion differs from findings based on data from earlier instruments. 
We are confident of our result based on the relatively high signal-to-noise of
the XMM-Newton spectra.  We note, however, that narrow, small equivalent width
absorption lines such as those inferred (at intermediate redshifts) from {\it
Chandra} grating observations of BL Lac objects by Nicastro et al. (2002; see
also Cagnoni et al. 2004), or those due to ionized oxygen in the hot halo of
our Galaxy (e.g., Nicastro et al. 2002, McKernan, Yaqoob \& Reynolds 2004), are
not excluded by the data in-hand.  The determination of presence or absence of
such features  require much higher S/N data than are available in our sample.

It is worth noting that a similar absence of strong spectral features was also
found in another,  independent study of the {\it XMM-Newton} X-ray spectra of
four of the five BL Lacs where {\it BBXRT} and {\it ASCA} spectra appeared to
show these line features, specifically H1219+301, Mkn 501, H1426+428 and
H0548-322 (Blustin et al. 2004).  Those authors reported that if any of the
previously found absorption features were indeed real, they had to represent a
transient phenomenon.  They found this particular conclusion unlikely, ruling
it out at 93\% confidence based on the assumption that the (multiple) {\it
XMM-Newton} observations of those objects represented random time instances. 
The objects in the Blustin et al. (2004) sample are not discussed here, and
moreover, our integration times are smaller so we have used primarily the EPIC
data and not the RGS.  However, our results are in the same vein.  With the
addition of our data, it appears unlikely that the  absorption features found
with {\it ASCA} and {\it BBXRT} were transient phenomena.

The most likely explanation of the previously reported line features is a
combination of  mis-calibration of the previous instruments and the use of
overly simplistic  spectral models.  As an example, a deficit of low energy
counts in the  {\it ASCA} data has been previously noted in reference to the
data for NGC 5548  (Iwasawa, Fabian, and Nandra 1999);  the simultaneous {\it
ASCA} and {\it ROSAT} PSPC  observation implied significantly higher flux below
0.5 keV than any  reasonable spectral extrapolation of the {\it ASCA} SIS data
would allow.   Another reason is that a spectral curvature of exactly the type 
discussed in this paper would, at low resolution and lower signal-to-noise,
mimic an absorption feature when the underlying spectrum was assumed to  be a
simple power law.  However, the confirmation that an absorption  feature was
measured would require data below the energy of such a  putative feature, and
such data were not always available (as was  the case for the {\it Einstein} 
SSS as well as SSS + MPC data).  We therefore suggest that the claims made by
earlier workers were a byproduct of  the particular simple power law model used,
while the effect was due to gradual spectral curvature  and/or instrumental 
effects rather than absorption features from material along the line of 
sight.  

%

\section{Modeling the X-ray and Broadband Spectral Characteristics}

We have constructed broadband spectral energy distributions (SEDs) for  each
object using the EPIC data, in conjunction with the data from the OM and the
UMRAO flux nearest in time to the {\it XMM} observation.  For objects which 
were too faint in either the optical or radio at the time of observation to
register a positive detection on either the OM or UMRAO, we have used POSS
plates to obtain an average optical flux and radio fluxes from the literature
(Perlman et al. 1996b and sources therein) for the SED.  We fit to these data a
simple parabolic model, in order to obtain a peak frequency, $\nu_{peak}$.   We
did not assume a standard synchrotron spectrum for this procedure because the
small amount of data we have do not adequately constrain these models (see
e.g., Leahy 1991, Pacholczyk 1970).     For a
few objects, where the initial fit was not good, we had to modify a single
non-simultaneous point to find a convincing value of $\nu_{peak}$.   

 In Table 6 we summarize the broadband spectral data and $\nu_{peak}$ for every
object in our sample. As can be seen, our objects range between log
$\nu_{peak}\approx 14.5-18$, as expected for HBL-type and intermediate objects. 
A few objects, noted in Table 6, have  very flat X-ray spectral index and/or
significant evidence  of $\nu_{peak} > 10^{18}$ Hz.  Our procedure yielded 
$\nu_{peak}$ values considerably in excess of  $10^{18}$ Hz for all these
objects (see also \S 3.1), but in Table 6 we list them as having $\nu_{peak} =
10^{18}$ Hz because our data are not  able to test for a higher peak frequency
given their frequency space coverage.

We then proceeded to test for correlations between broadband
spectral properties and both $\Gamma$ and $d\Gamma / d(log E)$.  We then
tested each for correlations using standard statistical tests (Spearman's rank
correlation, $\rho$, and Pearson's correlation coefficient, $r$).   These plots
and their physical implications are the main subject of this section. 

In Figure 5 we show plots of $\Gamma$ versus  $\alpha_{ro}$ (left),
$\alpha_{ox}$ (middle) and $\nu_{peak}$. As can be seen, our data show
significant anti-correlations between  $\Gamma$ and $\nu_{peak}$ 
($r=-0.741,\rho=-0.781, P=0.18$\%)   and between $\Gamma$ and $\alpha_{ox}$
($r=0.609, \rho=0.634, P=1.1$\%)  but no significant correlation between
$\Gamma$ and and $\alpha_{ro}$ ($r=0.036, \rho=-0.05, P=84.0$\%).  This
constellation of results is consistent with the $\alpha_x$ -- $\nu_{peak}$ 
relation of Padovani et al. (1997), when one considers that the range of
$\nu_{peak}$ values we cover is only $\sim 3$ decades.  

\subsection{Statistical Particle Acceleration?}

The observed spectra exhibit curvature  over a frequency range from $0.5 $ keV
to $10$ keV, in the sense of a gradual steepening with increasing energy.
Assuming that the observed emission is synchrotron radiation, this implies that
the electron distribution responsible for this emission is not the typical
power law predicted by  particle acceleration schemes (for a recent review see
Gallant 2002). Instead,  the electron spectrum has to gradually curve downwards
in a way that  produces the observed spectra. Given that X-ray synchrotron
emission requires {\it in situ} particle acceleration due to the short
radiative lifetimes of the emitting particles ($\tau_{sync} = 1.2 \times 10^3
B^{-3/2}_G E_{\rm keV}^{-1/2} \delta^{-1/2}$ s; Rybicki \&  Lightman 1979),
the ultimate physics behind curved X-ray spectra must be  connected on a very
basic level with the nature of particle acceleration in  these systems. 
Here and in the next subsection we discuss possible solutions to this
problem.

In Massaro et al. (2004b) a log-parabolic spectral curvature was fit to the
X-ray spectrum and broadband SED of Mkn~421 in several flux states.  Those
authors modeled the log-parabolic curvature via a statistical particle
acceleration process under which a particle $i$ with energy $\gamma_i$ in a
given region has a finite probability ($<1$) of being accelerated, where the
probability is given by $p_i =g/\gamma_i^q$, where $g, q$ are positive
constants.  They show in their paper that this leads analytically to a
log-parabolic curvature form (see their \S 6 and equations (10)-(18)), and 
predicts a linear correlation between the X-ray spectral index, $\Gamma_x$, and
$d\Gamma / d(log E)$. In Figure 6, we show plots of $\nu_{peak}$ versus
$d\Gamma / d(log E)$ (top) as well as  $\Gamma_x$ versus  $d\Gamma / d(log E)$
(bottom).  As can be seen, neither  of these plots shows a significant
correlation  ($\nu_{peak}$ versus  $d\Gamma / d(log E): r=0.615, \rho = 0.393, 
P=11.6$\% ;  $\Gamma_x$ versus  $d\Gamma / d(log E): r=-0.24, \rho = 0.014,
P=95.3$\%).   The results are similar for tests of $\nu_{peak}$ and $\Gamma$ 
versus $\beta$.  Thus our data do not support the model of Massaro et al.
(2004b).   One possible reason why this is so is that Massaro et al. (2004b)
model particle acceleration only, and do not include losses.  Moreover, the
statistical particle acceleration is still assumed to be time-invariant at all
locations in the jet and hence the issue of variability is not addressed.  

\subsection{Spectral Curvature: a  Signature of Episodic Particle
 Acceleration?}

The results of \S 4.1 prompt us to look for another explanation that takes
into account the known observational properties of jets and particularly
blazars.  Another motivation for this is the recent result of Perlman \& Wilson
(2005) that standard, continuous injection models of particle acceleration
consistently overpredict the X-ray flux of the M87 jet by large factors (up to
$\sim 100$ at 1 keV, varying with energy roughly as $E^{0.4}$).  The
interpretation advanced in that paper was one of a position- and
energy-variable  (but not time-variable) filling factor for particle
acceleration.  The Perlman \& Wilson work, however, does not explore the issue
of spectral curvature, as the statistics in most components of the M87 jet are
inadequate for this purpose; nor does it explore issues related to variability,
which are known to be  of paramount importance for BL Lac objects. 


In this subsection we address these issues directly, by demonstrating that 
curved electron particle spectra can be produced if the particle acceleration
is episodic.  Importantly, this model assumes only a time-variable particle
acceleration which in an integrated sense should be indistinguishable from a a
spatial average.   Unlike the case of M87, here we have no spatial
information regarding the distribution of jet X-ray emission and/or particle
acceleration.   Instead, we know that variability is an important
characteristic of blazar jets and thus we optimize our model for this
case. We explore these subjects more deeply in a later paper (Georganopoulos
\& Perlman, in preparation) that fully outlines a model considered in brief
form below.

Consider a zone, possibly a shock, where electrons with Lorentz factor
$\gamma_0$ are injected at a rate of $Q$ electrons per second 
and  accelerated to higher energies.
Following Kirk, Rieger, \& Mastichiadis (1998), the kinetic equation
describing particle acceleration is 

\begin{equation}
{ \partial n \over \partial t}={\partial \over \partial \gamma} 
\left[ \left({\gamma \over t_{acc}}-\beta \gamma^2 \right)n\right]+ {n \over t_{esc}}=Q \delta(\gamma-\gamma_0),
\end{equation}

{\noindent{where $n(\gamma,t)$ is the electron energy distribution,  
$1/t_{acc}$  is   the particle acceleration rate,  $1/t_{esc}$ is the escape
rate of particles from the acceleration region, and }}

\begin{equation}
\beta={3\over 4}{\sigma_T \over m_e c} U
\end{equation}

{\noindent{with $\sigma_T$ the Thomson cross-section and U the total energy
density of the magnetic field and ambient photons. Assuming that the injection
started at time $t=0$,  the electron energy distribution after time $t$ will be
a power law with electron index $s=1+t_{acc}/t_{esc}$ up to a Lorentz factor }}

\begin{equation}
\gamma_1(t)=\left( {1 \over \gamma_{max}}+                                      
\left[{1\over \gamma_0}-{1\over \gamma_{max}}\right ] e^{-t/t_{acc}}
 \right)^{-1},\label{gamma1}
\end{equation}

{\noindent{where $\gamma_{max}=1/\beta t_{acc}$ is the maximum energy
electrons  reach asymptotically  if  the particle acceleration mechanism
operates  for $t\gg \ln [(\gamma_{max}-\gamma_0)/\gamma_0]t_{acc}$. The time
$t$ required for electrons to be accelerated up to Lorentz factor $\gamma_1$
is }}

\begin{equation}
t(\gamma_1)=t_{acc}
\ln{\gamma_1(\gamma_{max}-\gamma_0)\over\gamma_0(\gamma_{max}-\gamma_1)}.
\label{time}
\end{equation}

If we assume that particle acceleration is characterized by a 
typical timescale $T$, 
then the electron distribution at any time 
$0\leq t\leq T $ is a power law that cuts off at  a Lorentz factor
$\gamma(t)$, reaching its maximum Lorentz factor $\gamma_1(T)$
 at $t=T$. Particle acceleration ceases
to operate at $t=T$, and  the total time $\tau$ over which
electrons of Lorentz factor $\gamma$ are provided by the particle
acceleration mechanism is

\begin{equation}
\tau=T-t=t_{acc}\ln{\gamma_1 (\gamma_{max}-\gamma) \over \gamma(\gamma_{max}-\gamma_1)}.
\label{eq:gamata}
\end{equation}

{\noindent{The time $\tau$ is also the time during which  an electron of
Lorentz factor $\gamma$  radiates, as long as the radiative cooling timescale 
$t_{rad}=1/\beta\gamma=t_{acc}\gamma_{max}/\gamma$ satisfies the condition
$t_{rad}\ll \tau$.  In other words,  $\tau$ is the total time the accelerated
electron distribution is non-zero at electron Lorentz factor $\gamma$; note
that, as is apparent from Eq. (5), the condition $t_{rad}\ll \tau$  can always
be satisfied for $\gamma_1$  sufficiently close to $\gamma_{max}$.  If the
light crossing time  of the emission region or the integration time  of our
observations is greater than $T$, then we effectively observe a  time averaged
electron distribution $\langle n(\gamma) \rangle $ which is the product  of a
power law  multiplied by a logarithmic term representing  the time interval 
this power law is available at each energy:}}

\begin{equation}
\langle n(\gamma)\rangle \propto\gamma^{-s}\ln{\gamma_1 (\gamma_{max}-\gamma) \over \gamma(\gamma_{max}-\gamma_1)},\;\gamma\leq\gamma_1.
\end{equation} 

{\noindent{ Using the the $\delta$-function approximation (i.e., an electron of
Lorentz factor $\gamma$ produces synchrotron photons  only at  the critical
synchrotron  energy) to calculate  the synchrotron  emission, we obtain an
observed photon spectrum}}

\begin{equation}
{dN \over dE} \propto E^{-(s+1)/2}\ln{E_1^{1/2} (E_{max}^{1/2}-E^{1/2})          \over E^{1/2}(E_{max}^{1/2}-E_1^{1/2})},\;E\leq E_1,
\end{equation} 

{\noindent{where $E_1$ and $E_{max}$ are the synchrotron energies produced by
electrons of Lorentz factor $\gamma_1$ and $\gamma_{max}$ respectively.  In
Figure 7 we show the photon spectrum and photon index as a function of energy,
for the following parameter choices: $s=3.0, E_{max}=100$ keV and $E_1=50$keV. 
As can be seen, the spectrum steepens gradually, with $\Gamma $ increasing at
a  fairly constant rate up to $\sim$ few keV, and then dramatically at
$E\gtrsim 10$ KeV. If our interpretation of the steepening is correct, we
expect to see an increasing curvature in future observations above the  $10$
keV upper limit of {\it XMM-Newton}, as predicted in equation (7) (see Figure
7).  One might also expect that in other objects, e.g., LBL, this cutoff
might occur at lower energies.  In LBL, however, we do not observe such steep
values of $\Gamma$, but rather, flat X-ray spectra are seen ($\Gamma<2$; see
Padovani et al. 2001, 2004), which are typically interpreted as  being
dominated by inverse-Comtpon emission (Padovani et al. 1997, Lamer et al.
1996). In practice, it may be difficult to observe this steepening because as
the curvature of the synchrotron component increases, the spectrum is
essentially cutting off and at some point the observed spectrum will begin to
be dominated by the onset of the Compton component.  What one might expect to
see in practice, then, would be concave spectra (i.e., increasing $\Gamma$) up
until the two components cross, with an inflection point at their crossing
point and then a constant value of $\Gamma$ ($<2$) at higher energies.  One
final thing that should be noted here is  that above 10 keV, almost all the HBL
spectra that currently exist are for objects that are currently in a flaring
state, where we would not expect to see significant steepening because the
particle acceleration is still at or near its maximum.  Also, there are no
currently available instruments that would be sufficiently sensitive at $E>10$
keV to search for further steepening in the spectrum of HBLs.}

The two energies $E_1$ and $E_{max}$ appearing in Equation (7)
can be derived through spectral fitting. However, their values
alone are offering little insight to the physical
conditions in the source,  because they  depend on  a series of
parameters  ($B$, $t_{acc}$, $\gamma_0$, $T$). As we discuss in \S 5,
additional information from frequency-dependent variability can be used to
constrain the physical state of the source.

\section{Conclusions}

We have presented 17 high-signal-to-noise {\it XMM-Newton} X-ray spectra for 13
BL Lac objects. These spectra do not show any spectral lines, contrary to
expectations from some previous observations.   We believe the reason for this
discrepancy is due to a combination of overly simplistic spectral models and/or
insufficiently precise calibration of previous instruments.  As a result the
intrinsic X-ray spectra of BL Lac objects  appear to be quite featureless,
giving no sign as to the physics of any region within the central engine, other
than the jet itself.  This is quite different from all other non-blazar type
sources, where some line features, in either  emission or absorption, are
seen.  This can be interpreted in the context of  unified scheme models (e.g.,
Urry \& Padovani 1995), under which the jet emission is Doppler boosted because
it is seen at small angles to our line  of sight.  It is, however, difficult to
address the issue of the necessary viewing angles to produce the degree of
featurelessness seen in these spectra, particularly given that at present our
knowledge concerning the X-ray spectra of the parent population of FR 1 radio
galaxies is limited.  Future work will provide much information on this
subject, but currently the best information available (based on luminosity
function estimates) suggests viewing angles $\theta \sim 10-30 $ and Lorentz
factors $\Gamma \sim 5$.  This would  yield Doppler boosts to the jet's
apparent luminosity of hundreds to thousands, as detailed in e.g., Urry \&
Padovani (1995).

Our data are best fit by curved spectra which grow steeper in an approximately
logarithmic fashion across the {\it XMM-Newton} band.  From an observational
point of view, a logarithmic curvature is not a new concept, having been first
introduced by Landau et al. (1986) for radio-optical observations of blazars,
and then revived for X-ray spectra by Giommi et al. (2002).  Our analysis is,
however, the first to demonstrate that intrinsic curvature is to be  preferred
over the alternate model of additional absorbing material along the line of
sight. From the theoretical point of view, logarithmically curved spectra
require explanation, as most of the standard models of synchrotron emission
produce spectra of a constant power-law slope (see, e.g., Leahy 1991 for a
review).   Massaro et al. (2004b) suggested statistical acceleration,  
whereby a given particle in the distribution has a probability of being
accelerated, as one possible model for this type of spectral shape.   The
predictions of that model are not borne out by our analysis, however, and a
statistical  model also does not take into account the variable nature of BL
Lacs.   It was with this motivation that we considered an episodic model for
particle acceleration,  which explicitly considers a modification to the
kinetic equation and thus includes both losses and a model of variability.  We
showed that this model fits our data reasonably well and predicts continued
steepening at higher energies, until the onset of the IC component.  

Episodic particle acceleration may be related to the broadband variability
observed  in TeV sources like 1959+650 (Krawczynski et al. 2004), where  both
the power and peak frequency of the synchrotron spectrum in X-rays  exhibit an
increase by more than a factor of $\sim 10$, while the optical flux remains
practically constant. This can be easily accommodated in our scheme through an
increase of  the characteristic time  $T$ that  particle  acceleration
operates.       In this case $\gamma_1$ and $E_1$ will both increase, resulting
in a more powerful  X-ray synchrotron spectrum that peaks at higher energies. 
The situation will be quite different at lower frequencies;    the curvature of
the spectrum produced by the logarithmic term will be present  down to photon
energies for which the  radiative cooling times $t_{rad}$ becomes larger than
the characteristic particle acceleration  time $T$. Below this energy the
electron energy distribution and the corresponding synchrotron spectrum will 
be a simple power law. At these lower energies an increase in $T$ can only 
affect the amplitude of the observed spectrum and even that only to the extent
that such an  increase results in an increased fraction of time that particle 
acceleration is operating (often called the duty cycle). If, for example, $T$
describes both the time  acceleration is on and off, then as $T$ increases  the
average number of particles injected remains constant, resulting in a constant
flux at lower frequencies, while the spectrum changes dramatically at high
frequencies due to the increase of both  $\gamma_1$ and $E_1$.  It is very
encouraging that our observations of 1ES1959+650, one to three months after
flare observed by Krawczynski et al. (2004), show significant  spectral
curvature.  This indicates that the characteristic timescales on which episodic
acceleration might operate is $\lsim 1$ month, which is consistent with the
findings of long-look campaigns (Perlman et al. 1999, Tanihata et al. 2001). 
Such a timescale is also consistent with Doppler-boosted versions of the flare
found recently in the jet  of M87 (Harris et al. 2003, Perlman et al. 2003),
requiring only values $\delta \sim 10$ (as opposed to the more modest
$\delta=1-2$ required for M87).  Indeed, the fact that we see logarithmic
curvature in 14 of 17 of these spectra indicates that the typical ratio of
$\tau/T \lsim 1;$ i.e., the amount of time in which flares are seen is not
appreciably smaller than the amount of time in which more gentle or no
variability is seen.  This is not inconsistent with the data from the ASCA
long-look campaigns (Tanihata et al. 2001).

\begin{acknowledgments}

ESP, MG and TD acknowledge support from NASA LTSA grant NAG5-9997 and
NASA XMM grant NAG5-10109.
GM and KA acknowledge support from NASA XMM grants NAG5-11972 and NAG5-13356, 
as well as the Department of Energy contract to SLAC no. DE-AC3-76SF00515.  
JHK would like to thank the Institute of Astronomy, Cambridge for their 
hospitality while part of this work was done, and the Raymond and
Beverly Sackler Fund for support during his visit there.  His work was
supported by NSF grants AST-0205806 and AST-0313031.  JTS acknowledges
support from NASA XMM grant NAG5-11890.  MFA acknowledges support from
NASA XMM grant NAG5-11973.

\end{acknowledgments}

\begin{figure}
\epsscale{0.9}
\plotone{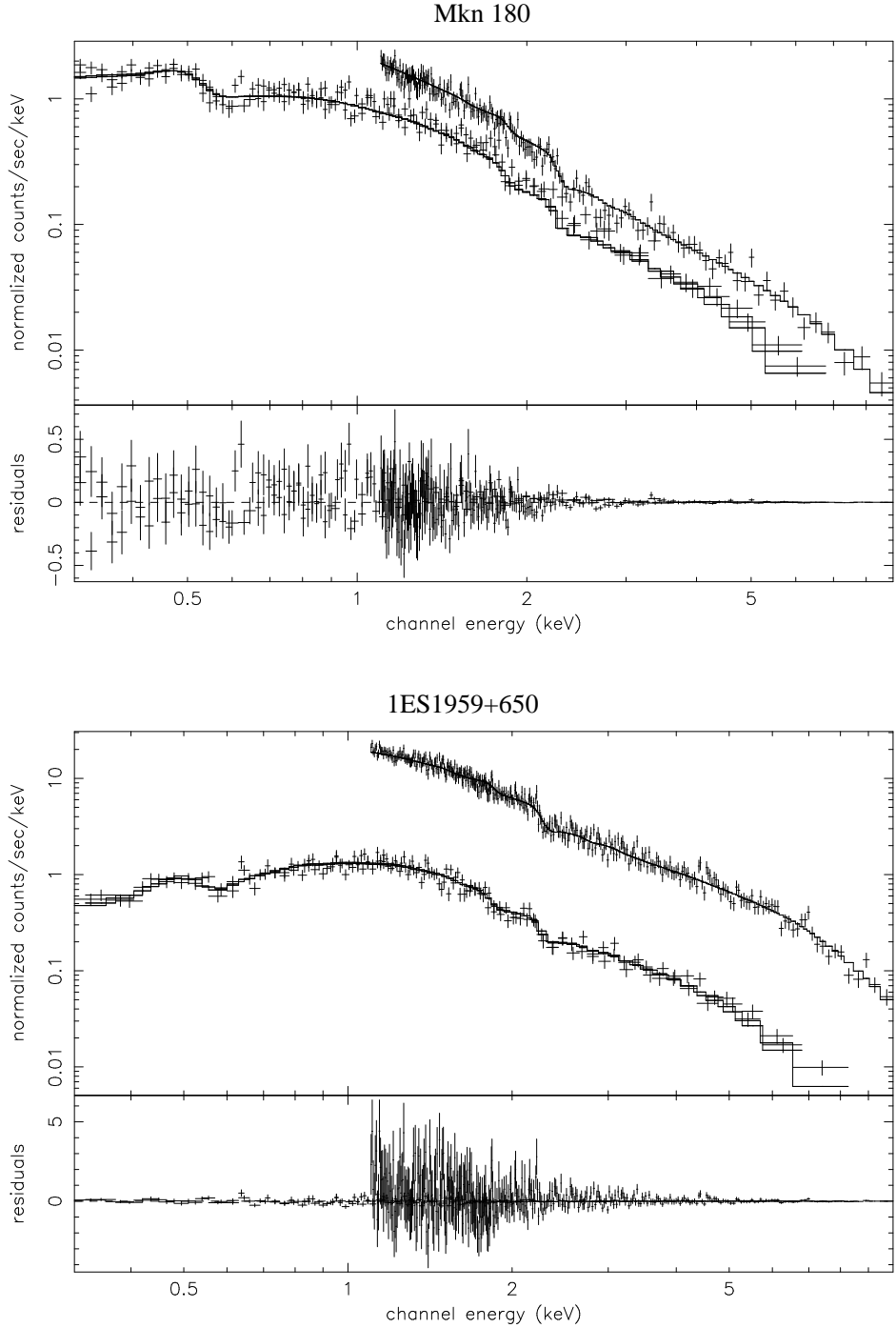}
\caption{Two examples of the XMM EPIC spectra.  At top, Mkn 180
(OBSID 0094170101) and at bottom, 1ES1959+650 (OBSID 0094383301). 
In both panels, 
the EPN data are the top line and the data from the MOS detectors are below.
The model shown represents the power law plus variable $N_H$ fit.
See Sections 2 and 3 for discussion.}

\end{figure}

\begin{figure}
\epsscale{0.95}
\plotone{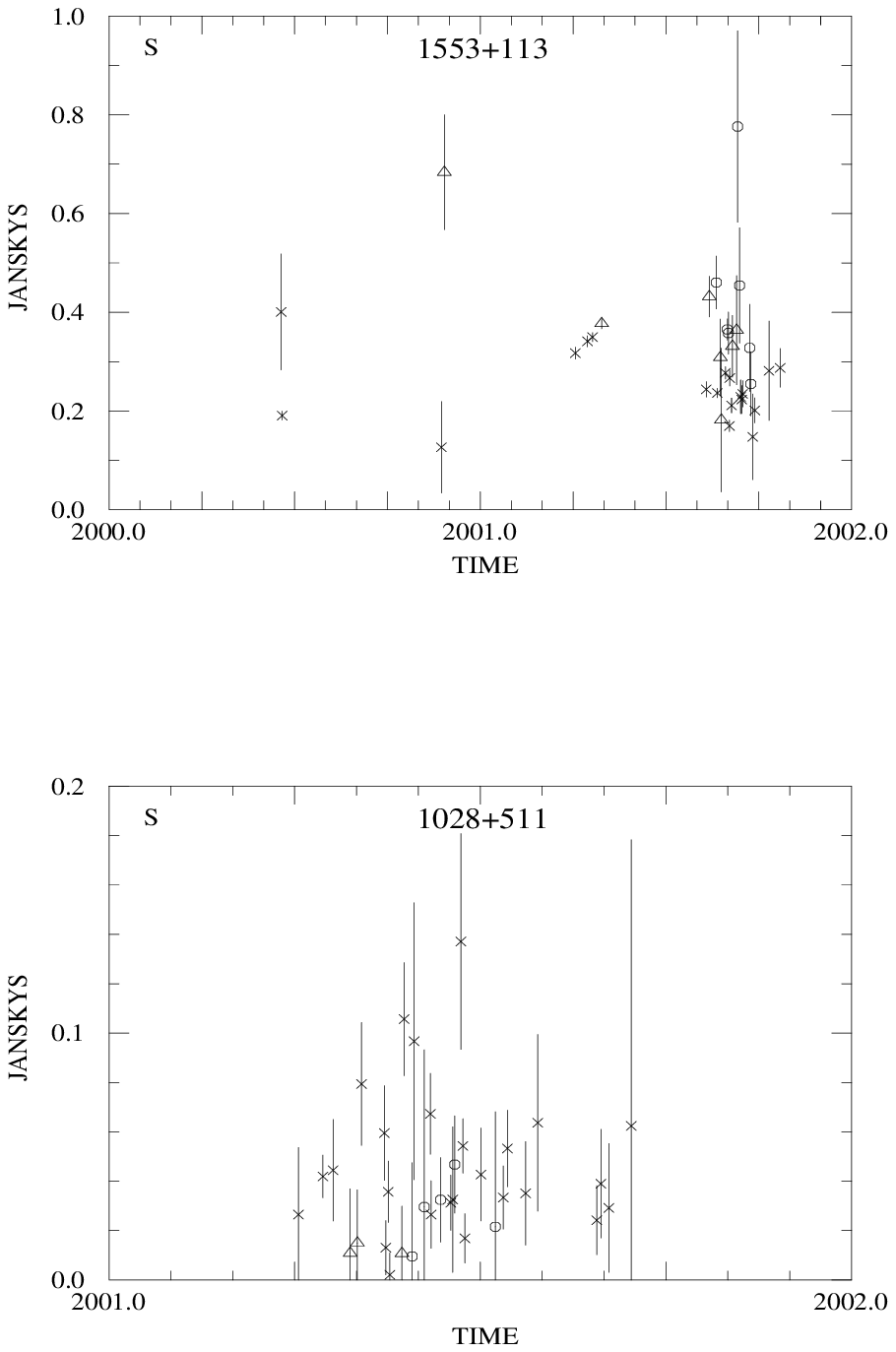}
\caption{Radio light curves for 1ES 1553+113 (top) and 1ES 1028+511 (bottom),
obtained at the University of Michigan Radio Astronomy Observatory.  
Data at 14.5, 8.0 and 4.8 GHz are denoted by crosses, circles and triangles
respectively.  Daily averages of the data are shown, with the bars denoting
$1\sigma$ error estimates.  As seen, both sources vary considerably during the
radio observations. See Section 2.1-2.2 for discussion.}

\end{figure}

\begin{figure}
\epsscale{1.10}
\plotone{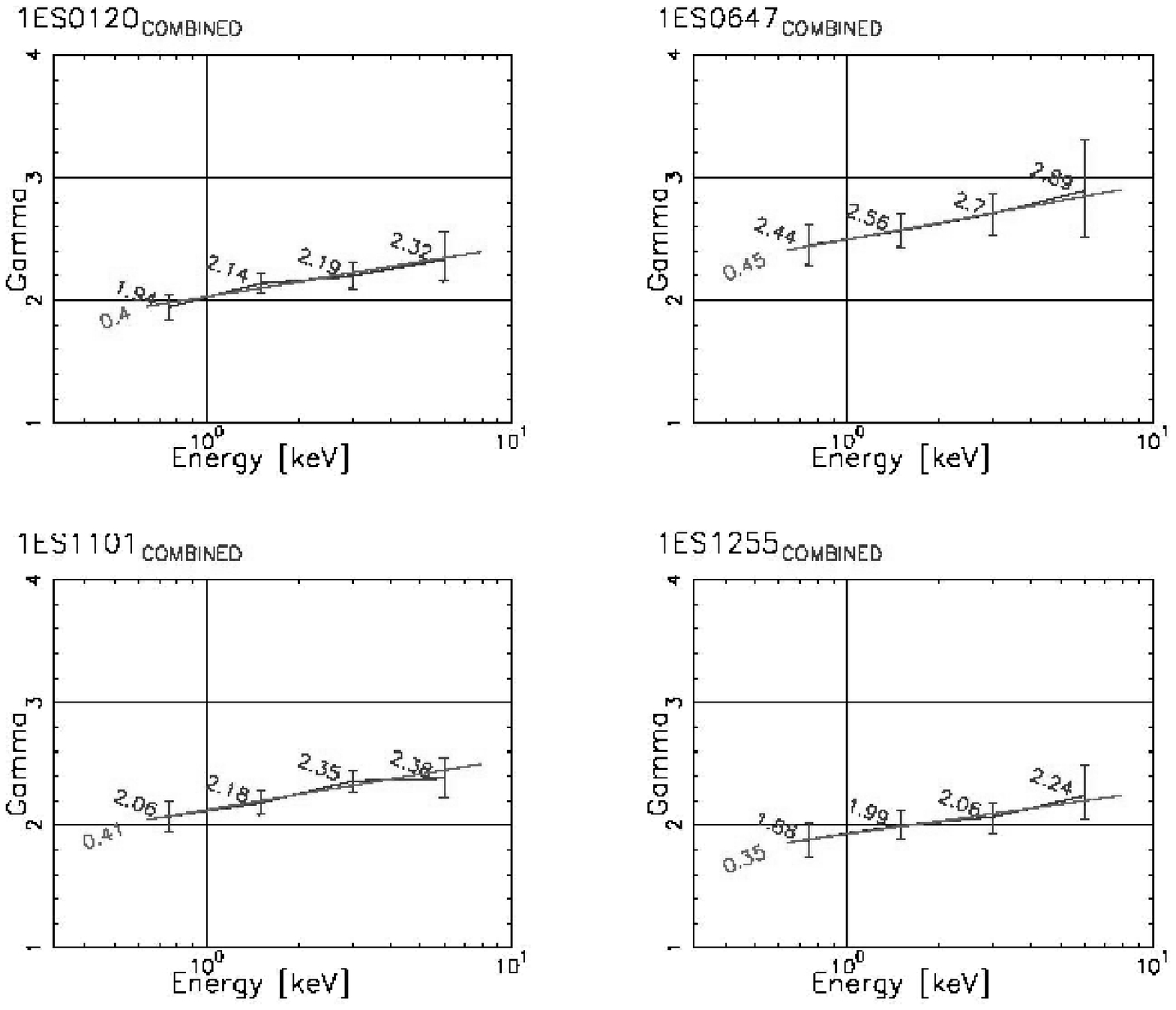}
\caption{Four examples of fitting the spectra of BL Lacs in our sample to
smaller spectral ranges.  All four of these objects show steeper spectra at
higher energies, with $d\Gamma / d(log E) \approx 0.4$.  See Section 3.3 for
discussion.}

\end{figure}

\begin{figure}
\epsscale{1.2}
\centerline{\plotone{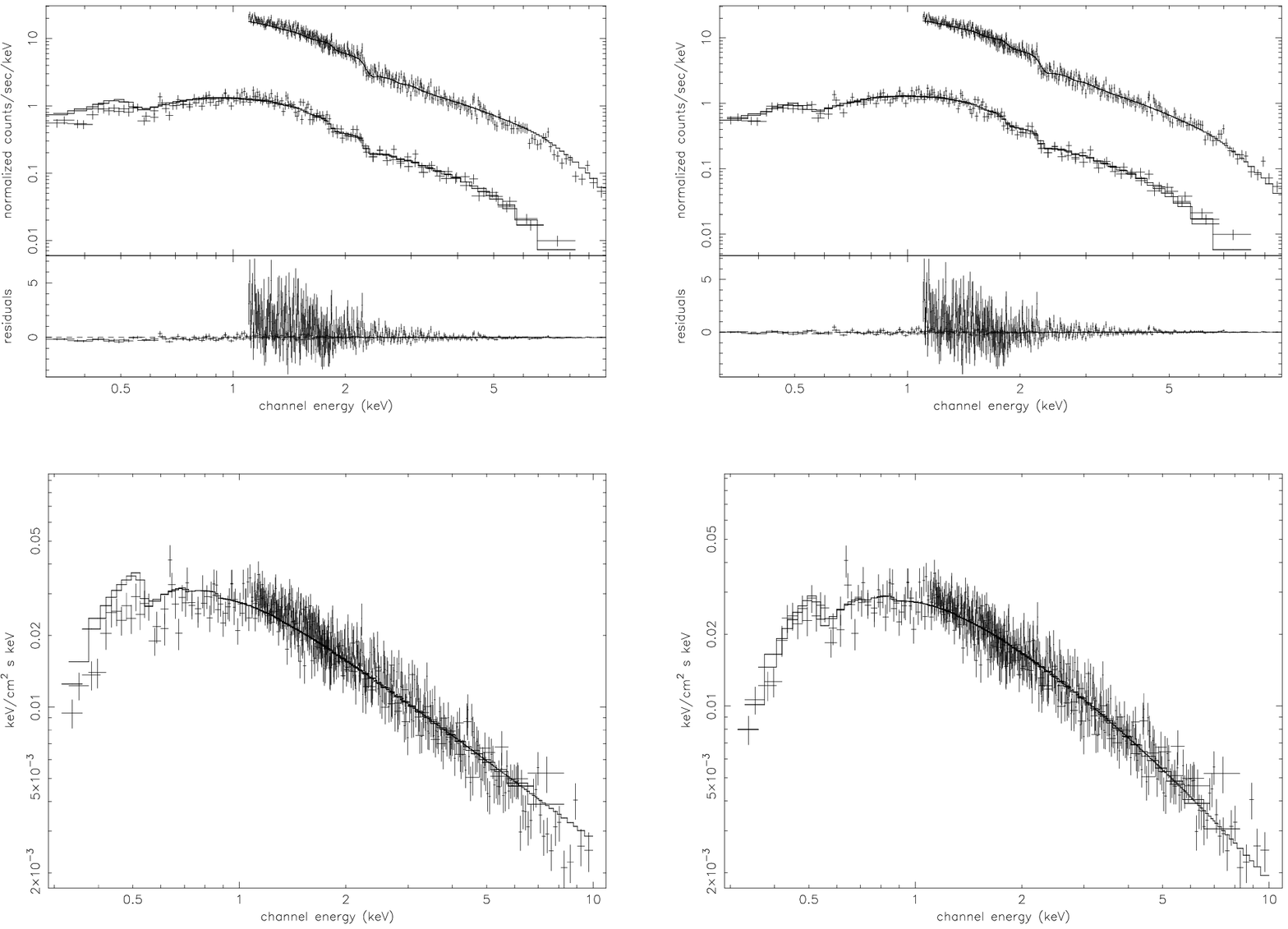}}
\caption{Comparison of model fits for the XMM spectrum of 1ES1959+650 
(OBSID 0094383301).  In the panel at left, the data were fitted with a 
power-law plus Galactic N(H) model, while at right, the data were fitted with 
the log-parabola plus Galactic N(H) model.  Note that the fit in the left 
panel is adequate between 1.5--5 keV, but cannot account for the curvature seen 
both at energies $<1.5$ keV and $>5$ keV.  At top, the comparison is shown in
terms of the detected spectrum, with each plot showing received spectra plus
folded model for the MOS 1 \& 2 (bottom lines) and PN (top lines), as well as
residual. At bottom, we show the unfolded energy spectrum.  These plots show
that curvature is  present in this spectrum up to the highest energies seen by
XMM.}

\end{figure}

\begin{figure}
\epsscale{1.25}

\centerline{\plotone{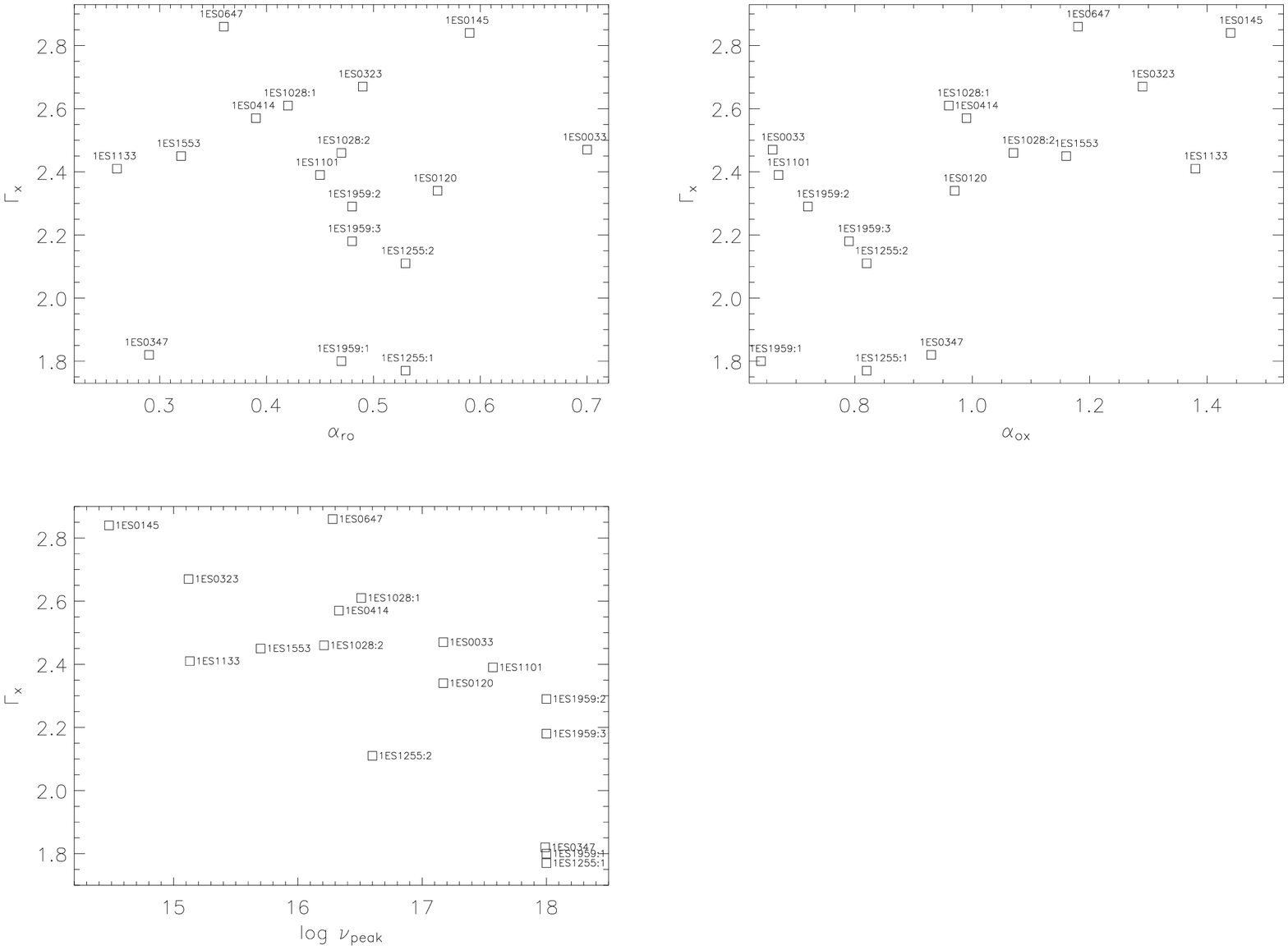}}

\caption{Plots of the X-ray spectral slope $\Gamma$, versus broadband spectral
index $\alpha_{ro}$ (top left) and $\alpha_{ro}$ (top right), as well as the 
SED peak frequency $\nu_{peak}$ (bottom left).  For easy cross-referencing 
with the Tables, we note in these plots the identity of each point. See Section
4.1 for discussion.} \end{figure}

\begin{figure}
\epsscale{0.9}
\plotone{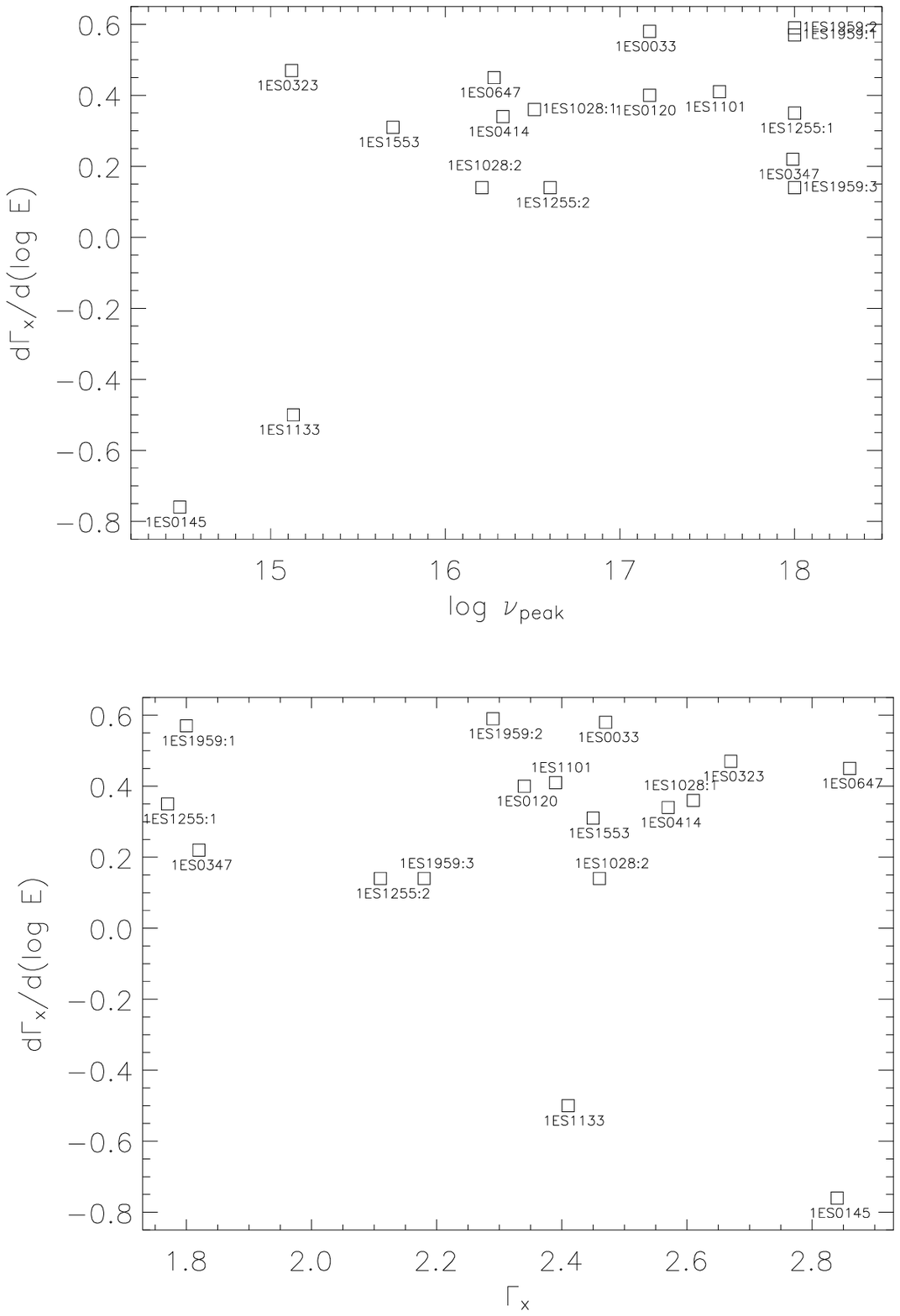}
\caption{Plots of $d\Gamma/d(logE)$ versus $\Gamma$ (top) and $\nu_{peak}$
(bottom). For easy cross-referencing 
with the Tables, we note in these plots the identity of each point.
See Section 4.1 for discussion.}
\end{figure}

\begin{figure}
\epsscale{0.8}
\plotone{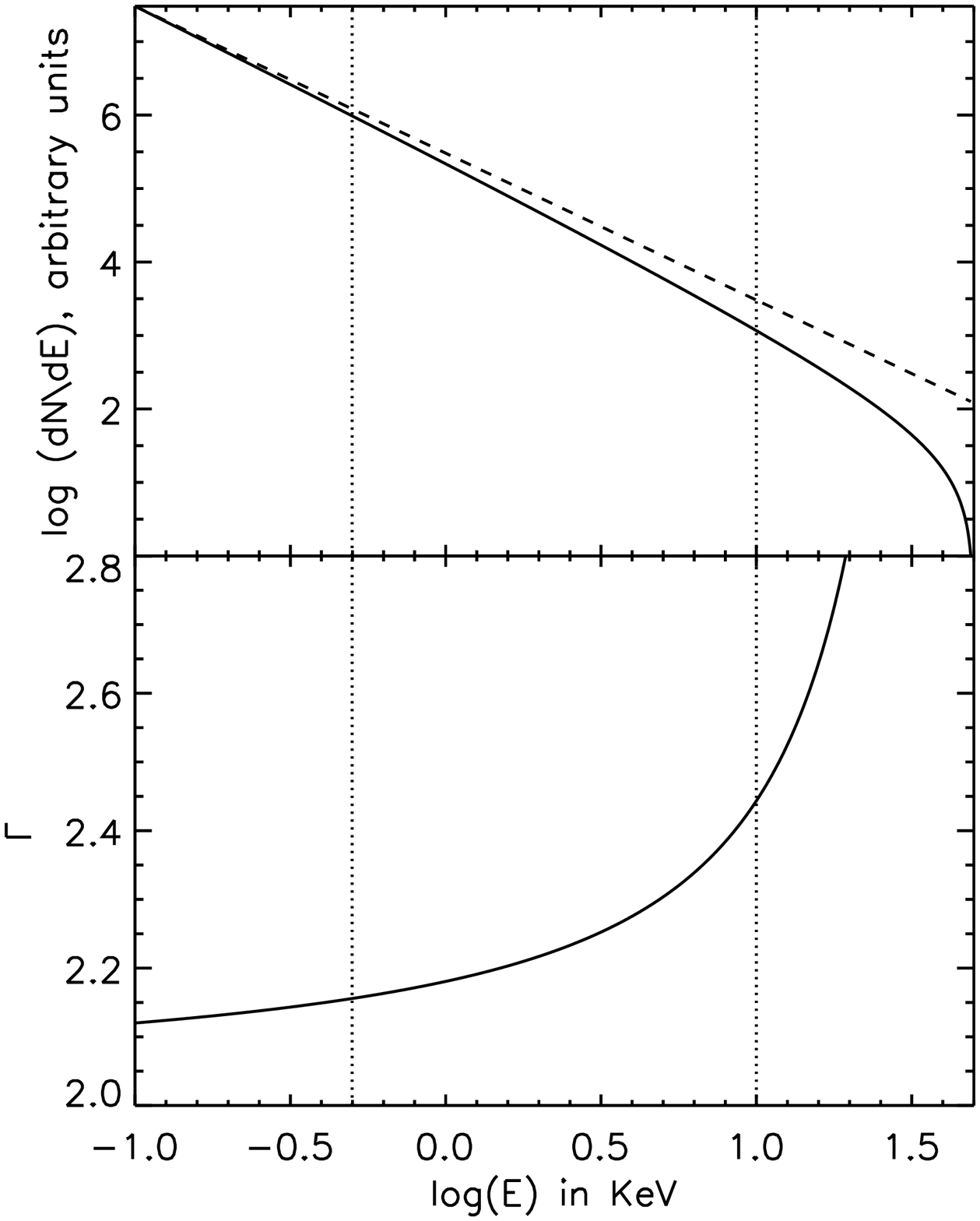}
\caption{Upper panel: The photon spectrum (solid line) 
resulting from an episodic  injection of electrons in the particle
acceleration process. Here $s=3.0$, $E_{max}=100$ KeV, and $E_1=50$ KeV. 
Note that $E_1$ is the maximum emitted energy the source is reaching in this 
episodic mode of injection, while $E_{max}$ is the emitted energy the 
source would reach under steady  injection conditions.
 For comparison,  a power law emission corresponding to an electron 
distribution with $s=3.0$ is also shown (broken line). Lower panel: the 
photon index $\Gamma$ that corresponds to the emitted photon spectrum. 
Note the gradual steepening of the spectrum, manifested by an increase of 
$\Gamma$ with energy. The dotted lines represent the energy range $0.5-10$ keV 
covered by {\it XMM}.
See Section 4.2 for discussion. }
\label{theory}
\end{figure}

\begin{landscape}
\begin{deluxetable*}{cllcccccccccc} 
\tablecolumns{12}
\tablewidth{0pt}
\tablecaption{Log of XMM Observations}

\tablehead{
\colhead{1ES Name}&\colhead{OBSID}&\colhead{Date}&    
\colhead{MOS gt}&\colhead{Filter} &\colhead{Frame}& \colhead{Ct Rate}&\colhead{PN gt}& \colhead{Filter}& \colhead{Frame}&
\colhead{Ct Rate} \\
&&&\colhead{(sec)}&&& \colhead {(sec)}&\colhead{(sec)} &&& \colhead{(Ct/s)}
}

\startdata

0033+595&	0094381301$^1$&	2003-02-01&	1500&	 Thin/Med & PartialW2& 1.17&	4100&	Thin1 & Small &3.43\\
0120+340&	0094382101&	2002-01-05&	5600&	Medium & PartialW2 &2.43&	3500&	Medium & Small &8.48\\
0145+138&	0094383401&	2002-07-18&	4200&	Thin1 & PartialW2 &0.07&	3200&	Thin1 & Small &0.25\\
0323+022&	0094382501&	2002-02-05&	5200& Medium & PartialW2 &	0.81&	3200&	Medium & Small &2.81\\
0347$-$121&	0094381101&	2002-08-28&	3000&	 Thin1 & PartialW2 &3.32&	2200&	Thin1 & Small & 11.6 \\
0414+009&	0094383101&	2002-08-26&	5700&Thin1 & Partial/ Full &	3.18&	5600&	Thin1 & Small &12.1\\
0647+250&	0094380901$^{1,2}$&2002-03-25&	2300&	Medium & PartialW2 &2.10&	1900&Medium & Small &	7.49\\
... 	&	0094382901&	2002-03-25&	N/A&	 N/A & N/A & N/A &	N/A&	Medium & Small &N/A\\
1028+511&	0094381801$^1$&	2001-05-15&	N/A & N/A& N/A&	N/A &1176&	 Medium & Small&	 5.59\\
... 	&	0094382701&	2001-11-26&	5600&Medium & PartialW2 &	4.30&	3500&	Medium & Small &16.2\\
1101$-$232&	0094380601$^1$&	2001-05-29&	2200&	 Medium & PartialW2 & 4.70&	4300&	Medium & Small &17.6\\
1133+704&	0094170301$^1$&2001-04-12&	N/A&	 Thin1 & Full &N/A&	N/A&	Thin1 & Full & N/A 	 \\
...	&	0094170101$^2$&	2001-04-12&	3300& Thin1 & Full &	4.45 (2.02) $^3$&	6900&Thin1 & Full &	12.5 (5.81)$^3$\\ 
1255+244&	0094383001$^1$&	2002-12-12&	5800&	Medium & PartialW2 &1.30&	4000&	Medium & Small &4.40\\
...	&	0094383201&	2002-06-26&	662&	 Medium & PartialW2 &1.35&	435&	Medium & Small &4.88\\
1553+113&	0094380801&	2001-09-06&	 5000 &Medium & PartialW2 & 	13.0 (4.31) $^3$&	3500&	Medium & Small &50.5\\
1959+650&	0094380201$^1$&	2002-11-23&	300&	 Medium & PartialW2 & 21.2 (5.29)$^3$ &	N/A&	Medium & Small &N/A\\
... 	&	0094383301$^1$&	2003-01-16&	2200& Medium & PartialW2 & 	10.5 (2.08)$^3$ &	1200&	Medium & Small &39.5\\
...	&	0094373501&	2003-02-09&	3700&	 Medium & PartialW2 & 12.5 (3.80) $^3$&	2000&Medium & Small &	43.3\\

\enddata
\tablenotetext{1}{This observation was affected by proton flaring}
\tablenotetext{2}{Observation was attempted again in the same orbit}
\tablenotetext{3}{Count Rates are raw count rates in the effective exposure interval, while values in parentheses are rates after pile-up correction.}
\end{deluxetable*}
\clearpage
\end{landscape}

\begin{deluxetable*}{ccccc} 
\tablecolumns{5}
\tablewidth{0pt}
\tablecaption{Supporting observations}
\tablehead{
\colhead{1ES Name}&\colhead{OBSID}&\colhead{OM Band}&\colhead{F(UV)} & \colhead{$F_{max}$(rad) }\\
&&&\colhead{($10^{-16} {\rm ~erg~cm^{-2}~s^{-1} ~\AA^{-1}}$)} &\colhead{(Jy)}
}
\startdata
0033+595&	0094381301&	UVW2$\times$2 & $<3.3^1$ & 0.3 \\	
0120+340&	0094382101&	UVW2$\times$5 & $8.9 \pm 0.9$  & 0.1	\\	
0145+138&	0094383401&	UVW2$\times$4 & $<3.3^1$ & 0.2	\\	
0323+022&	0094382501&	UVW2$\times$5 & $14.4 \pm 1.0$ & 0.3	\\	
0347$-$121&	0094381101& UVW2$\times$4 & $7.6 \pm 2.0$ & ---$^2$\\	
0414+009&	0094383101&	UVW2$\times$5 & $18.5 \pm 0.1$ & 0.1	\\	
0647+250&	0094380901&	UVW2$\times$1 & $<3.3^1$& 0.1\\
...	&	0094382901&	UVW2$\times$5 & $<3.3^1$& 0.1 	\\	
1028+511&	0094381801&UVW2$\times$5 & $18.2 \pm 1.6$ & 0.2 \\	
...	&	0094382701&	UVW2$\times$4 & $21.9 \pm 0.7$ & 0.2	\\
1101$-$232&	0094380601&	UVW2$\times$4 & $10 \pm 3$ & 0.3\\	
1133+704&	0094170301&	N/A &	   N/A  & ---$^2$ \\ 
...	&	0094170101&	U$\times$4 & $272 \pm 8 $ & N/A$^3$	 \\ 	
1255+244&	0094383001&	UVW2$\times$5 & $3.2 \pm 1.5$ & 0.2 \\	
...	&	0094383201&	UVW2$\times$4 & $<3.3^1$ & 0.2  \\	
1553+113&	0094380801&	UVW2$\times$5 & $185 \pm 5$ & 0.8	\\	
1959+650&	0094380201&	UVW2$\times$5 & $35.0 \pm 2.0$ & 0.6 \\	
...	&	0094383301&	UVW2$\times$5 & $25.3 \pm 2.0$ & 0.6 \\	
...	&	0094373501&	UVW2$\times$5 & $28.4 \pm 2.0$ & 0.6	\\	
\enddata
\tablenotetext{1}{Object not detected in the OM observation; value is a $2 \sigma$ upper limit}
\tablenotetext{2}{Below detection threshold in all UMRAO observations}
\tablenotetext{3}{No UMRAO observations made as this object was originally part of another program}

\end{deluxetable*}

\clearpage

\begin{landscape}

\begin{deluxetable*}{cccccccccccccc}
\tablecaption{BL Lac Spectral Fit Parameters}
\tablewidth{0pt}
\tablehead{
\colhead{1ES Name} & 
\colhead{ObsId} &
\multicolumn{4}{c}{Fit parameters, N(H) Fixed} & 
\multicolumn{4}{c}{Fit parameters, N(H) Free } & \\
& & 
\colhead{$\mathrm{N_{H, Gal}}$} &  
\colhead{$\Gamma$} &
\colhead{$\mathrm{F_{2-10 keV}}$} &
\colhead{$\chi^2$/dof. (dof.)} &
\colhead{$\mathrm{N_{H, tot}}$} &
\colhead{$\Gamma$} &
\colhead{$\mathrm{F_{2-10 keV}}$} &
\colhead{$\chi^2$/dof. (dof.)} & 
\colhead{$F$-test}\\
& & 
\colhead{$10^{20}~\mathrm{cm^{-2}}$} & & 
\colhead{erg cm$^{-2}$ s$^{-1}$} & & 
\colhead{$10^{20}~\mathrm{cm^{-2}}$} & &
\colhead{erg cm$^{-2}$ s$^{-1}$} & &
\colhead{Probability}
 } 
\startdata
0033+595 & 0094381301\tablenotemark{F}	& 42.4 & $2.2^{+0.02}_{-0.03}$ & $1.2\times10^{-11}$ & 1.42(288)  & 63.8$^{+3.8}_{-3.6}$ & 2.47$\pm{0.05}$ & 1.1$\times10^{-11}$ & 1.01(287) & $<10^{-10}$ \\
0120+340 & 0094382101 		      	& 5.14 & $2.16^{+0.01}_{-0.01}$& $9.7\times10^{-12}$ & 1.70(527)  & 9.9$\pm{0.6}$  & 2.34$^{+0.02}_{-0.03}$ & 8.9$\times10^{-12}$ & 1.32(526) & $<10^{-10}$\\
0145+138 & 0094383401 		      	& 5.10 & $2.07\pm 0.12$        & $2.5\times10^{-13}$ & 1.44(88)   & 22.1$^{+8.0}_{-6.9}$ & 2.83$^{+0.38}_{-0.34}$ & 1.6$\times10^{-13}$ & 1.24(87)  & $3.2\times10^{-4}$\\
0323+022 & 0094382501 			& 8.74 & $2.54^{+0.02}_{-0.03}$& $2.0\times10^{-12}$ & 1.05(213)  & 11.9$\pm{1.2}$ & 2.67$\pm{0.06}$ & 1.9$\times10^{-12}$ & 0.95(212) & $1.3\times10^{-5}$ \\
0347$-$121 & 0094381101 		& 3.64 & $1.77^{+0.02}_{-0.01}$& $1.8\times10^{-11}$ & 1.18(530)  & 5.0$\pm{0.6}$  & 1.82$\pm{0.03}$ & 1.8$\times10^{-11}$ & 1.16(529) & $2.6\times10^{-3}$ \\
0414+009 & 0094383101 			& 10.3 & $2.47^{+0.02}_{-0.01}$& $9.3\times10^{-12}$ & 1.20(499)  & 13.1$\pm{0.6}$ & 2.57$\pm{0.03}$ & 8.9$\times10^{-12}$ & 1.09(498) & $<10^{-10}$\\
0647+250 & 0094380901 			& 12.8 & $2.62\pm 0.02$        & $5.6\times10^{-12}$ & 1.36(278)  & 19.0$^{+1.2}_{-1.1}$ & 2.86$\pm{0.05}$ & 5.0$\times10^{-12}$ & 1.05(277) & $<10^{-10}$\\
1028+511 & 0094381801\tablenotemark{F}  & 1.16 & $2.44^{+0.04}_{-0.03}$& $1.2\times10^{-11}$ & 1.20(107)  & 3.77$\pm{1.1}$ & 2.61$\pm{0.08}$ & 1.0$\times10^{-11}$ & 1.04(106) & $<10^{-10}$\\
...      & 0094382701 			& 1.16 & $2.28\pm 0.01$        & $1.2\times10^{-11}$ & 1.41(600)  &  5.2$\pm{0.4}$ & 2.46$\pm{0.02}$ & 9.7$\times10^{-12}$ & 0.98(599) & $<10^{-10}$ \\
1101$-$232 & 0094380601\tablenotemark{F}& 5.76 & $2.23^{+0.02}_{-0.01}$& $2.3\times10^{-11}$ & 1.37(682)  & 10.9$\pm{0.6}$ & 2.40$^{+0.02}_{-0.03}$ & 2.2$\times10^{-11}$ & 1.05(681) & $<10^{-10}$ \\
1133+704 & 0094170101			& 1.42 & $2.40\pm 0.02$        & $9.8\times10^{-12}$ & 1.21(361)  & 1.8$\pm{0.7}$  & 2.42$\pm{0.04}$ & 9.7$\times10^{-12}$ & 1.20(360) & 8.4 $\times10^{-2}$ \\
1255+244 & 0094383001\tablenotemark{F}  & 1.26 & $1.99^{+0.01}_{-0.02}$& $5.1\times10^{-12}$ & 1.18(380)  & 2.9$^{+2.9}_{-2.6}$ & 1.83$\pm{0.12}$ & 4.8$\times10^{-12}$  & 1.02(379) & $<10^{-10}$  \\
...      & 0094383201 			& 1.26 & $1.76^{+0.07}_{-0.06}$& $7.5\times10^{-12}$ & 0.97(134)  & 4.6$\pm{0.8}$ & 2.11$^{+0.04}_{-0.03}$ & 4.8$\times10^{-12}$  & 0.97(133)  & N/A \\
1553+513 & 0094380801 			& 3.67 & $2.38\pm 0.01$        & $3.5\times10^{-11}$ & 1.26(815)  & 5.7$\pm{0.4}$  & 2.46$^{+0.01}_{-0.02}$ & 3.4$\times10^{-11}$ & 1.17(814)  & $<10^{-10}$ \\ 
1959+650 & 0094380201\tablenotemark{F,a}& 10.1 & $1.70\pm 0.05$        & $2.9\times10^{-10}$ & 1.44(58)   & 13.3$^{+3.1}_{-2.9}$ & 1.80$^{0.1}_{-0.11}$ & 2.8$\times10^{-10}$ & 1.41(57)  & 0.275 \\
...      & 0094383301\tablenotemark{F}  & 10.1 & $2.10^{+0.02}_{-0.01}$& $7.7\times10^{-11}$ & 1.51(514)  & 17.9$\pm{1.0}$ & 2.29$^{+0.02}_{-0.03}$ & 7.2$\times10^{-11}$ & 1.08(513) & $<10^{-10}$  \\
...      & 0094383501 			& 10.1 & $2.02\pm 0.02$        & $6.9\times10^{-11}$ & 1.64(863)  &16.9$\pm{0.6}$ & 2.18$^{+0.02}_{-0.01}$ & 6.5$\times10^{-11}$  & 1.13(862)& $<10^{-10}$ \\

\enddata

\tablenotetext{a}{No PN data available, fit parameters are for MOS data only.}
\tablenotetext{F} {Most of this observation is contaminated by proton flaring.}
\tablecomments{MOS data used in 0.5 - 10.0 keV range PN in 1.1 - 10.0 keV range.}
\end{deluxetable*}
\clearpage
\end{landscape}

\begin{landscape}

\begin{deluxetable*}{ccccccccccc}
\tablecolumns{11}
\tablewidth{0pt}
\tablecaption{Parameters for BL Lac Spectral Curvature}
\tablewidth{0pt}
\tablehead{
\multicolumn{2}{c}{}&\multicolumn{6}{c}{4-band $\Gamma$-Fitting$^a$} & \multicolumn{3}{c}{Logarithmic Parabola$^a$}
\\
\colhead{1ES Name} & 
\colhead{ObsId} &
\colhead{$\Gamma_{0.5-1.0 keV}$} &
\colhead{$\Gamma_{1.0-2.0 keV}$} &
\colhead{$\Gamma_{2.0-4.0 keV}$} &
\colhead{$\Gamma_{4.0-10.0 keV}$} & 
\colhead{$\chi2$/dof. (dof.)}$^b$ &
\colhead{$\mathrm{d}\Gamma / \mathrm{d}(log E)$} &
\colhead{$\Gamma$} & \colhead{$\beta$} & \colhead{$\chi2$/dof. (dof.)} 
}

\startdata

0033+595 & 0094381301 & $1.49^{+1.07}_{-1.98}$ & $2.03^{+0.18}_{-0.18}$ &$2.34^{+0.12}_{-0.12}$ & $2.33^{+0.22}_{-0.22}$ & $1.04(187)$& $0.58$ & $1.71^{0.1}_{-0.09} $&$-0.67^{+0.12}_{-0.13}$  & 1.10 (287)   \\
0120+340 & 0094382101 & $1.94^{+0.1}_{-0.1}$ &    $2.14^{+0.08}_{-0.08}$ &$2.19^{+0.1}_{-0.09}$ &   $2.32^{+0.22}_{-0.16}$ & $1.16(383)$  &$0.4$ &  $2.05\pm 0.02$  & $-0.30\pm 0.04$ & 1.37 (526)  \\
0145+138 & 0094383401 & $0.25^{+0.86}_{-3.25}$ & $2.83^{+0.72}_{-0.66}$ &$2.54^{+0.91}_{-0.71}$ & $0.16^{+1.12}_{-2.76}$ & $1.34(77)$ & $-0.76$  & $1.84^{+0.17}_{-0.2}$ & $-1.26^{+0.56}_{-0.68}$ & 1.27(87) \\
0323+022 & 0094382501 & $2.29^{+0.17}_{-0.17}$ & $2.6^{+0.15}_{-0.15}$ &  $2.52^{+0.18}_{-0.19}$ & $2.89^{+0.53}_{-0.38}$ & $1.03(156)$  &$0.47$ & $2.48^{+0.03}_{-0.04}$ &$-0.22^{+0.09}_{-0.08}$ & 0.96(212)   \\
0347$-$121 & 0094381101 & $1.6^{+0.12}_{-0.13}$ &   $1.82^{+0.09}_{-0.09}$ &$1.9^{+0.1}_{-0.11}$ &    $1.75^{+0.13}_{-0.17}$ & $1.15(398)$  &$0.22$ & $1.75^{+0.03}_{-0.02}$ & $-0.05^{+0.05}_{-0.04}$ & 1.18(529)   \\
0414+009 & 0094383101 & $2.4^{+0.11}_{-0.11}$ &   $2.45^{+0.09}_{-0.09}$ &$2.51^{+0.09}_{-0.09}$ & $2.82^{+0.18}_{-0.18}$ & $1.15(309)$  &$0.34$ & $2.41 \pm 0.02$ & $-0.16\pm 0.04$ & 1.11(498)    \\
0647+250 & 0094380901 & $2.44^{+0.16}_{-0.16}$ & $2.56^{+0.13}_{-0.13}$ &$2.7^{+0.17}_{-0.18}$ &   $2.89^{+0.4}_{-0.38}$ &  $0.89(197)$  &$0.45$ & $2.5^{+0.03}_{-0.04}$ & $-0.39 \pm 0.08$ & 1.10(277)     \\
1028+511 & 0094382701 & $2.23^{+0.07}_{-0.07}$ & $2.23^{+0.07}_{-0.06}$ &$2.49^{+0.09}_{-0.09}$ & $2.48^{+0.17}_{-0.17}$ & $0.93(425)$  &$0.36$ & $2.20 \pm 0.01$ &  $-0.30 \pm 0.03$ &          1.00(599)    \\
...	 & 0094381801$^c$ & ---  & ---  & --- & --- & --- &   ---   & ---    $2.42  \pm 0.04$ &         $-0.19^{+0.10}_{-0.12}$ & 1.13(106)   \\
1101$-$232& 0094380601& $2.06^{+0.12}_{-0.12}$ & $2.18^{+0.09}_{-0.09}$ &$2.35^{+0.08}_{-0.08}$ & $2.38^{+0.16}_{-0.15}$ & $0.93(472)$  &$0.41$ & $2.09 \pm 0.02$ & $-0.32^{+0.03}_{-0.04}$ & 1.05(681)   \\
1133+704 & 0094170101 & $2.3^{+0.14}_{-0.14}$ &   $2.21^{+0.14}_{-0.14}$ & $2.34^{+0.14}_{-0.14}$ & $2.47^{+0.28}_{-0.27}$ & $1.28(223)$&$0.14$ & $2.39^{+0.03}_{-0.02}$& $-0.01^{+0.05}_{-0.06}$  & 1.21 (360) \\
1255+244 & 0094383001 & $1.88\pm{0.13}$ &  $1.99\pm0.11$ & $2.06^{+0.12}_{-0.13}$ & $2.24^{+0.25}_{-0.18}$ & $0.97(283)$  & $-0.5$ &  $1.90^{+0.03}_{-0.02}$ & $-0.22\pm 0.05$&  1.05(379)  \\
...	 & 0094383201 & $2.29^{+0.48}_{-0.49}$ & $1.70\pm0.37$ & $1.71^{+0.41}_{-0.45}$ & $1.08^{+0.58}_{-0.71}$ & $0.84(112)$  &$0.35$ & $1.76^{+0.10}_{-0.10}$ & $-0.01 ^{+0.18}_{-0.21}$ &  0.98(133)    \\
1553+113 & 0094380801 & $2.35^{+0.07}_{-0.07}$ & $2.25^{+0.07}_{-0.07}$ & $2.42^{+0.06}_{-0.06}$ & $2.44^{+0.12}_{-0.12}$ & $1.11(603)$  &$0.14$  & $2.33^{+0.02}_{-0.01}$ &$-0.13^{+0.02}_{-0.03}$&  1.17(814)  \\
1959+650 & 0094380201 & $1.34^{+0.33}_{-0.35}$ & $1.78^{+0.25}_{-0.25}$ & $1.87^{+0.32}_{-0.31}$ & $1.29^{+0.6}_{-0.6}$ &     $1.45(57)$  & $0.31$ & $1.64\pm{0.10}$ & $-0.14^{+0.18}_{-0.19}$ &  1.43(57)    \\
...	 & 0094383301 & $1.75^{+0.19}_{-0.19}$ & $2.14^{+0.12}_{-0.12}$ & $2.21^{+0.09}_{-0.09}$ & $2.36^{+0.16}_{-0.16}$ & $1.11(331)$  &$0.57$ & $1.88^{+0.04}_{-0.03}$ & $-0.39 \pm 0.05$ &  1.14(513)    \\
...	 & 0094383501 & $1.86^{+0.11}_{-0.11}$ & $1.89^{+0.07}_{-0.07}$ & $2.11^{+0.06}_{-0.06}$ & $2.36^{+0.1}_{-0.1}$ &    $1.05(653)$  &$0.59$ & $1.81\pm{0.02}$  & $-0.37^{+0.03}_{-0.04}$ &  1.13(862)  \\

\enddata
\tablenotetext{a}{$N_H$ assumed fixed at Galactic (see Table 3) for all fits.}
\tablenotetext{b}{$\chi^2$/dof. figures were derived by separately adding the $\chi^2$ and dof figures and only then dividing.}
\tablenotetext{c}{Four-band not performed because only PN data of low S/N were available}

\end{deluxetable*}
\clearpage
\end{landscape}

\begin{landscape}

\begin{deluxetable*}{llcccl} 
\tablecolumns{6}
\tablewidth{0pt}
\tablecaption{Discrete features found in RGS Spectra}

\tablehead{
\colhead{1ES Name$^a$}&\colhead{OBSID} & \colhead{Energy (keV)} & \colhead{$\Delta \chi^2$} & \colhead{Width (eV)} & \colhead{Comments }}

\startdata

1ES0033+595   & 0094381301 & 1.07 & $-12.02$ & $45^{+20}_{-24}$ & MOS 2 + 1 RGS only \\
1ES1101$-$232 & 0094380601 & 0.67 & $-18.56$ & $7^{+4}_{-3}$    & RGS 2 only \\
1ES1101$-$232 & 0094380601 & 1.36 & $-16.62$ & $22 \pm 9$       & RGS 2 only \\
1ES1255$+$244 & 0094383201 & 1.29 & $-11.98$ & $30 \pm 16$      & MOS 1, MOS 2 but not in RGS \\
1ES1255$+$244 & 0094383201 & 2.91 & $-8.91$  & $8^{+6}_{-4}$    & Single MOS 1\&2 channel, instrumental? \\
1ES1959$+$650 & 0094383301 & 1.92 & $-11.77$ & $<18$            & Two RGS 1 channels, not RGS 2 or MOS \\

\enddata
\tablenotetext{a}{This table lists only spectra where discrete features were found.  The spectra where the search
was performed but no features were found are listed in \S 3.3}

\end{deluxetable*}
\clearpage
\end{landscape}

\begin{landscape}

\begin{deluxetable*}{cccccl} 
\tablecolumns{6}
\tablewidth{0pt}
\tablecaption{Multiwaveband Parameters}

\tablehead{
\colhead{1ES Name}&\colhead{OBSID} & \colhead{$\alpha_{RO}$} & \colhead{$\alpha_{OX}$}&\colhead{log $\nu_{peak}$} & \colhead{Comments}
}

\startdata

0033+595&	0094381301&	0.70 &	0.66	& 17.17 &  Converges poorly; radio flux had to be lowered artificially \\
0120+340&	0094382101&	0.56 &	0.97	& 17.17 &      \\
0145+138&	0094383401&	0.59 &	1.44	& 14.48 & \\
0323+022&	0094382501&	0.49 &	1.29	& 15.12 &  Confusing radio source?  Catalog flux 25\% of UMRAO value \\
0347$-$121&	0094381101&	0.29 &	0.93	& 17.99 & $\alpha_x<1$ -- use log $\nu_{peak}$=18. \\
0414+009&	0094383101&	0.39 & 	0.99	& 16.33 & \\
0647+250&	0094380901&	0.36 & 	1.18	& 16.28 & Catalog optical flux 15x brighter than OM data \\
0647+250&	0094382901&	0.36 & 	1.18	& 16.28 & Catalog optical flux 15x brighter than OM data \\
1028+511&	0094381801&	0.42 &	0.96	& 16.51 & \\
...	&	0094382701&	0.47 & 	1.07	& 16.21 & \\
1101$-$232&	0094380601&	0.45 &	0.67	& 17.57 & \\
1133+704&	0094170301&	0.26 &	1.38	& 15.13 & \\
1255+244&	0094383001&	0.53 &	0.82	& 16.60 & Catalog opt flux 25x brighter than OM; increased opt flux $2\times$\\
...	&	0094383201&	0.53 &	0.82	& 18.00 & $\alpha_x<1$ -- use log $\nu_{peak}=18$.  \\
1553+113&	0094380801&	0.32 &	1.16	& 15.70 & \\
1959+650&	0094380201&	0.47 & 	0.63	& 18.00 & $\alpha_x<1$ -- use log $\nu_{peak}=18$.  \\
...	&	0094383301&	0.48 & 	0.72	& 18.00 & Used log $\nu_{peak}=18$ even though object not in flare. \\
...	&	0094373501&	0.48 &	0.79	& 18.00 & Used log $\nu_{peak}=18$ even though object not in flare. \\
\enddata
\end{deluxetable*}
\clearpage
\end{landscape}

\end{document}